# Interference Management based on RT/nRT Traffic Classification for FFR-Aided Small Cell/Macrocell Heterogeneous Networks

Mostafa Zaman Chowdhury, *Member, IEEE*, Md. Tanvir Hossan, *Student Member, IEEE* and Yeong Min Jang, *Member, IEEE*
Dept. of Electronics Engineering, Kookmin University, Seoul 136-702, Korea
E-mail: mzaman@kookmin.ac.kr, mthossan@ieee.org, yjang@kookmin.ac.kr

*Abstract*—Cellular networks are constantly lagging in terms of the bandwidth needed to support the growing high data rate demands. The system needs to efficiently allocate its frequency spectrum such that the spectrum utilization can be maximized while ensuring the quality of service (QoS) level. Owing to the coexistence of different types of traffic (e.g., real-time (RT) and non-real-time (nRT)) and different types of networks (e.g., small cell and macrocell), ensuring the QoS level for different types of users becomes a challenging issue in wireless networks. Fractional frequency reuse (FFR) is an effective approach for increasing spectrum utilization and reducing interference effects in orthogonal frequency division multiple access (OFDMA) networks. In this paper, we propose a new FFR scheme in which bandwidth allocation is based on RT/nRT traffic classification. We consider the coexistence of small cells and macrocells. After applying FFR technique in macrocells, the remaining frequency bands are efficiently allocated among the small cells overlaid by a macrocell. In our proposed scheme, total frequency-band allocations for different macrocells are decided on the basis of the traffic intensity. The transmitted power levels for different frequency bands are controlled based on the level of interference from a nearby frequency band. Frequency bands with a lower level of interference are assigned to the RT traffic to ensure a higher QoS level for the RT traffic. RT traffic calls in macrocell networks are also given a higher priority compared to nRT traffic calls to ensure the low call-blocking rate. Performance analyses show significant improvement under the proposed scheme compared to conventional FFR schemes.

## I. INTRODUCTION

Researchers all over the world are investigating optimum solutions to support the fast-growing demands of cellular communication connectivity by ensuring maximum utilization of limited and valuable wireless resources as well as provisioning the high quality of service (QoS) levels that are demanded by different traffic classes. Coexistence of small cells and macrocells is a good approach for providing the connectivity required by growing traffic. Small cells with inexpensive and low-power base stations (BSs) divert a huge amount of cellular traffic from macrocell networks to small cell networks [1]-[5]. Moreover, orthogonal frequency division multiple access (OFDMA) technology [6]-[10] has been widely applied in existing and next-generation cellular networks to increase frequency utilization. Although we have small cells for offloading traffic from congested macrocells and OFDMA technology to increase the frequency utilization, the macrocell networks still need to handle the frequency spectrum effectively. Even though intracell interference is avoided by orthogonal subcarrier allocation between macrocell user equipment (MUE) in each of the macrocells in an OFDMA network, intercell interference (ICI) still exists when frequency bands are reallocated between different macrocells [6]. ICI limits the overall spectral efficiency of the network, particularly for users near the boundary of a macrocell. Fractional frequency reuse (FFR) [11]-[15] is an excellent approach for improving frequency utilization as it can mitigate ICI in multicell OFDMA networks. The FFR technique is well-suited to OFDMA-based macrocellular networks in which the cells are divided spatially into center and edge regions with different frequency reuse factors [16]-[20].

FFR in macrocells is also an efficient technique for allocating the remaining frequency bands among the small cells within an overlaid macrocell to maximize the overall frequency utilization. Various interference management schemes such as averaging, avoidance, and coordination have been proposed to mitigate ICI [21]-[23]. In this paper, we provide a novel FFR solution for allocating the frequency bands. Our scheme carefully handles the transmitted power of different frequency bands to reduce interference effects.

*A. Related Recent Works and Motivation*

FFR schemes are already being used as an intercell interference coordination (ICIC) technique in OFDM-based wireless standards such as IEEE 802.16 m [24] and 3GPP-LTE Release 8 and above [25] to improve the performance of the cellular networks. The features of frequency and power domains ICIC do not provide any substantial gain in heterogeneous networks (HetNets). To serve users at reduced power levels in Long Term Evolution-Advanced (LTE-A) HetNets, enhanced inter-cell interference coordination (eICIC) [26]-[30] and Further eICIC (FeICIC) have been proposed in 3GPP LTE Release-10 and Release-11, respectively. The eICIC deals with interference mitigation on traffic and control channels. This is applicable in time domain in addition to frequency and power domains. It protects low power downlink small cell transmissions by mitigating interference from high power macrocells [27]. Time domain eICIC is realized over the use of Almost Blank Subframes (ABS). ABSs are subframes with reduced transmit power on some physical channels. In addition, cell selection bias is applied in eICIC to compensate the received signal power from aggressor macrocellular BSs (MBSs) to a victim small cell user. The eICIC performs muting of certain subframes at the macro layer. The victim small cell utilizes the subframes muted by the aggressor macrocell. Due to muting of few subframes at the macro layer, the spectrum utilization is slightly reduced.

Recent research on FFR has focused on the optimal design of FFR systems by utilizing advanced techniques to maximize the network resource utilization. In [6], an adaptive rate scheduling for a transmitting node employing FFR is proposed regardless of whether it is a BS or a mobile user. This scheme did not care about RT an nRT traffic calls. The authors in [7] derive the optimal signal-to-interference-ratio thresholds to maximize the coverage probability for both FFR and soft frequency reuse (SFR) networks. In [8], a self-organized dynamic FFR scheme is proposed which dynamically allocates resources to cell inner and outer regions in LTE-A relay-based networks. This work did not consider dynamic power allocation for interference reduction. An FFR in ultra-dense-network millimetre wave at 26 GHz band is investigated in [9]. This paper mainly focuses on dense network with short inter site distance, and higher order sectorisation. In [12], the authors investigate the tradeoff between the downlink ergodic spectral efficiency and the energy efficiency of distributed antenna systems by using a mathematical approach. This scheme did not consider real-time (RT) and non-real-time (nRT) traffic calls for frequency and power allocation. The energy efficiency of downlink transmissions in HetNets employing FFR is investigated in [14]. They formulate the joint cell-center boundary selection for FFR, scheduling, and power allocation problems. In [15], the authors develop an analytical framework targeting the downlink performance evaluation of FFR-aided OFDMA based two-tier HetNets. In [16], the authors propose a solution for FFR based on the center of gravity of users in each sector. Their scheme enables a distributed and adaptive solution for interference coordination.

Motivated by the researches on FFR, in this work, we propose a novel FFR scheme based on traffic classification and controlling of transmitted power. An FFR scheme can provide good performance metrics (e.g., signal-to-interference-plus-noise ratio (SINR), spectral efficiency, outage probability, and system throughput). However, FFR schemes based on fixed amount of frequency-band allocation [10], [15], [16] for different macrocells can leave valuable parts of the spectrum unutilized. Improper power allocation for the center-zone frequency bands in the FFR scheme can also cause serious interference for center-zone users. Another issue is ensuring different QoS levels for different classes of traffic according to their demands. Center-zone users can be supported through multiple frequency-band options, with each frequency-band option causing a different level of interference. Hence, there is a scope to allocate these frequency bands between different classes of traffic based on the QoS requirement and priority of the traffic classes. The QoS requirements are not as stringent for nRT traffic as they are for RT traffic [31]. Thus, better-quality frequency bands (in terms of interference) can be assigned for RT traffic compared to the nRT traffic calls to ensure better QoS for RT traffic calls. Also, RT traffic calls can be given higher priority compared to nRT traffic calls in frequency band allocation. Thus, the call-blocking rate of RT traffic calls can be reduced.

### B. Contributions

In this paper, we propose a new FFR scheme for macrocells. We consider frequency-band reassignment, efficient power allocation for different frequency bands, and RT/nRT traffic classification. Effective reassignment of bandwidth increases the spectral efficiency and reduces interference. Proper power allocation for different frequency bands reduces the effects of interference. Frequency-band allocation based on RT/nRT traffic classification ensures the QoS level of RT traffic calls along with an increased QoS level of nRT traffic calls. We also allocate frequency bands for small cells in the center and edge zones of different macrocells. The small cells within a macrocell are allowed to use the frequency bands that are not assigned to the overlaid macrocell, thus reducing cross-tier interference and increases overall spectral utilization. Our main contributions in this paper can be summarized as:

- A new frequency allocation scheme is proposed considering small cell/macrocell networks, RT/nRT traffic classification, and macrocell center/edge zones to maximize the resource utilization along with ensuring QoS level for higher priority RT traffic calls.
- A new FFR scheme is proposed in which the bandwidth allocation is based on RT/nRT traffic classification.
- After applying FFR technique in macrocells, frequency bands are efficiently allocated among the small cells that are not used by the overlaid macrocell.
- The total allocated frequency band for a macrocell is made flexible, i.e., dynamic frequency bandwidth can be borrowed from others based on traffic load. Hence, the overall call-clocking is decreased and resource utilization is increased.
- The transmitted power levels for different frequency bands are controlled to reduce the interference level.
- Assured level of QoS is provided for the RT calls by allocating frequency bands with a lower level of interference without significant reduction of the QoS level for nRT traffic calls.

### C. Organization

The rest of this paper is organized as follows. Section II gives an overview of the interference and traffic scenarios in small cell/macrocell HetNets. Our proposed class-based frequency allocation is presented in Section III. In Section IV, interference analysis and the effects of the interference are shown in detail. The performance of our proposed model is evaluated in Section V. Finally, a conclusion is drawn in Section VI. For ease of reference, the various notations used in this paper are summarized in Table I.

TABLE I
LIST OF NOTATIONS

| Notation | Meaning | Notation | Meaning |
|---|---|---|---|
| $A, B, C$ | Frequency bands for edge zones of different macrocells | $I_{2+}$ | Received interference power from $2^{nd}$ and higher-tier macrocells |
| $Z$ | Frequency band for center zone of every macrocell | $R_{k,i(C)}$ | Achievable data rate of the $k^{th}$ MUE in the center zone of the $i^{th}$ macrocell |
| $S_s^C$ | Set of allowed frequency bands for small cells in center zone | $R_{k,i(E)}$ | Achievable data rate of the $k^{th}$ MUE in the edge zone of the $i^{th}$ macrocell |
| $S_s^E$ | Set of allowed frequency bands for small cells in edge zone | $N_C$ | Number of channels allocated to the center zone of a macrocell |
| $F_{s,i}^C$ | Allocated frequency band for a small cell situated in the center zone of the $i^{th}$ macrocell | $N_E$ | Number of channels allocated to the edge zone of a macrocell |
| $F_{s,i}^E$ | Allocated frequency band for a small cell situated in the edge zone of the $i^{th}$ macrocell | $N_0$ | Received noise power at an MUE |
| $P_{t,M}$ | Maximum transmitted power by an MBS | $ASE_{macro}$ | Net area spectral efficiency of a macrocell in [bits/sec/Hz/macrocell] |
| $P_{t,s}$ | Transmitted power by an sBS | $C_C$ | Capacity per unit frequency in bits/sec/Hz for the center zone |
| $\alpha$ | A coefficient that determines the transmitted power of the center zone | $C_E$ | Capacity per unit frequency in bits/sec/Hz for the edge zone |
| $X_A$ | A frequency sub-band of $A$ that is reassigned to macrocell 7 | $\Delta_C$ | Frequency reuse factor for the center zone |
| $X_B$ | A frequency sub-band of $B$ that is reassigned to macrocell 7 | $\Delta_E$ | Frequency reuse factor for the edge zone |
| $X$ | Total frequency band reassigned to macrocell 7 | $F_{7(s)}$ | Total allocated frequency bands for the small cells within macrocell 7 |
| $X_{A,I}$ | Interfering frequency band of $X_A$ | $F_{7(s,C)}$ | Total allocated frequency bands for the small cells within macrocell 7 center zone |
| $X_{B,I}$ | Interfering frequency band of $X_B$ | $F_{7(s,E)}$ | Total allocated frequency bands for the small cells within macrocell 7 edge zone |
| $X_{i,I}$ | Interfering frequency band of $i^{th}$ macrocell | $SINR_{n,i(C)}$ | The received SINR level for the $n^{th}$ sUE in the small cell within center zone of $i^{th}$ macrocell |
| $X_I$ | Total interfering frequency band of $X$ | $SINR_{n,i(E)}$ | The received SINR level for the $n^{th}$ sUE in the small cell within center zone of $i^{th}$ macrocell |
| $A_{i,O}$ (or $B_{i,O}$) | Occupied frequency band in the $i^{th}$ macrocell | $h_{n,i(C)}$ | Channel gain between the $n^{th}$ sUE and the center zone antenna of the $i^{th}$ MBS |
| $A_R$ (or $B_R$) | Reserve frequency band in the $i^{th}$ macrocell | $h_{n,i(E)}$ | Channel gain between the $n^{th}$ sUE and the edge zone antenna of the $i^{th}$ MBS |
| $A_{i,V}$ (or $B_{i,V}$) | Releasable frequency band in the $i^{th}$ macrocell | $L_{macro}$ | Path loss for macrocell |
| $C_7 (=C)$ | Allocated frequency band for the edge zone of macrocell 7 | $d$ | Distance between the MBS and the user in kilometers |
| $BW_{req}$ | Required bandwidth for all calls | $f_c$ | Center frequency of the MBS antenna in MHz |
| $BW_{req,RT}$ | Required bandwidth for RT calls | $h_m$ | Height of user in meters |
| $BW_{req,nRT(E)}$ | Required bandwidth for nRT calls in the edge zone | $h_b$ | Height of MBS in meters |
| $BW_{req,nRT}$ | Required bandwidth for nRT calls | $L_{ow}$ | Penetration loss |
| $SINR_{k,i(C)}$ | Received SINR level for the $k^{th}$ MUE in the center zone of the $i^{th}$ macrocell | $L_{femto}$ | Path loss for small cell |
| $SINR_{k,i(E)}$ | Received SINR level for the $k^{th}$ MUE in the edge zone of the $i^{th}$ macrocell | $z$ | Distance between the sBS and the user in meters |
| $h_{k,i,(C)}$ | Channel gain between the $k^{th}$ MUE and center-zone antenna of the $j^{th}$ MBS | $f$ | Center frequency of the sBS antenna in MHz |
| $h_{k,i,(E)}$ | Channel gain between the $k^{th}$ MUE and edge-zone antenna of the $j^{th}$ MBS | $N$ | Distance power loss coefficient |
| $h_{k,s}$ | Channel gain between the MUE and sBS | $L_f$ | Floor penetration loss factor |
| $SINR_{k,7(C)}^{RT}$ | Received SINR level for the $k^{th}$ RT MUE in the center zone of macrocell 7 | $P_{outage,k,i}$ | Connection outage probability of the $k^{th}$ MUE in the $i^{th}$ macrocell |
| $SINR_{k,7(C)}^{nRT}$ | Received SINR level for the $k^{th}$ nRT MUE in the center zone of macrocell 7 | $\gamma$ | Threshold value of SINR |
| $\delta$ | A coefficient whose value is either 1 or 0 | $S_0$ | Received signal power by an MUE from the MBS |

## II. INTERFERENCE SCENARIOS IN HETNETS

In a heterogeneous network that comprises small cells and macrocells, the level of interference depends on the relative positions of the following four basic network entities: small cell BS (sBS), MBS, MUE, and small cell user equipment (sUE). There are four different link types (small cell downlink, small cell uplink, macrocell downlink, and macrocell uplink) that create unwanted interference, and may affect the other basic network entities [32]. Fig. 1 presents interference scenarios in a coexisting small cell/macrocell HetNets environment. An sUE receives interference from the downlinks of overlaid and neighboring macrocells if the MBSs and serving sBS are allocated the same frequency. The situation is of particular concern when the connected sBS is close to the MBS and the sUE is located at the edge of the small cell such that the transmitted high power from the MBS can possibly cause huge interference in the receiver of the sUE. An sUE is also affected by small cell downlinks of neighboring sBSs. The small cell downlinks affect the performance of nearby MUE. Therefore, small cell downlinks cause interference to nearby small cell user receivers as well as those of macrocell users. Whenever an MUE is close to a small cell coverage area, the uplink signal from the MUE to the connected MBS can cause interference with the receiver of an sBS. MUE can also cause interference with the neighboring MBSs. If a small cell is close to the MBS, the transmitted uplink signal from the sUE can cause interference with the receiver of MBS. Hence, co-tier as well as cross-tier interferences exist and may degrade the performance of different entities i.e., MBS, sBS, MUE, and sUE. Therefore, these interferences should be carefully handled to maximize frequency utilization and QoS level.

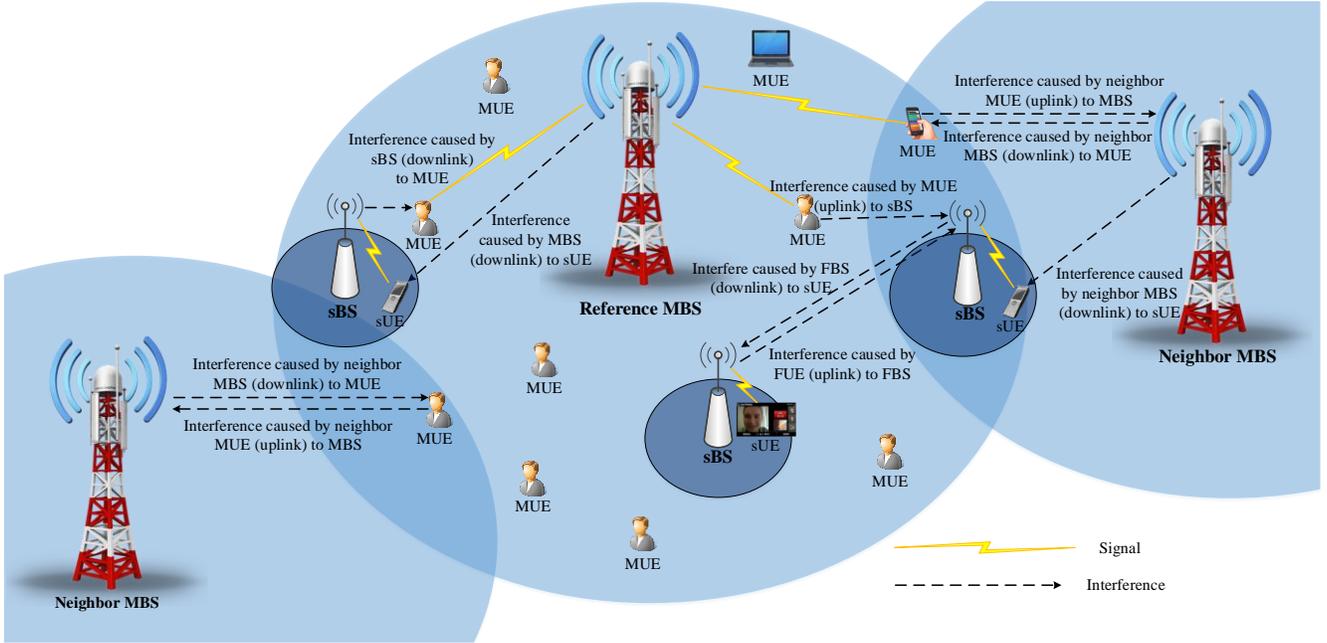

**Fig. 1.** Interference scenarios in a coexisting small cell/macrocell environment.

Currently, cellular networks provide wireless connectivity for diverse traffic types. Broadly, we can classify them as RT and nRT traffic. This traffic is supported by macrocells as well as small cells. A huge amount of RT/nRT traffic is distributed over the macrocell coverage area; in addition, thousands of small cells can be deployed in the macrocell coverage area. A careful frequency allocation based on RT traffic priority can ensure a better QoS level for RT traffic in terms of low call-blocking probability, higher SINR, and lower outage probability without significant reduction of the QoS level for nRT traffic calls.

## III. FREQUENCY ALLOCATION BASED ON FFR AND TRAFFIC CLASSIFICATION

FFR is a very well-known frequency-allocation technique for cellular communication. We propose a new frequency-allocation scheme for different macrocells in a cluster and the high density of small cells within the macrocells. Our proposed scheme is based on a frequency reuse factor of one and three for the center and edge zones of the macrocells, respectively. The frequencies of small cells are allocated in such a manner that frequency utilization is maximized without increasing interference. Fig. 2 shows the frequency allocation among three macrocells and small cells. Frequency band $Z$, a part of the whole band, is allocated to the center zones of all the macrocells. The remaining frequency band is equally divided into three sub-bands $A$, $B$, and $C$. Each of these bands are allocated to the edge zones of different macrocells of cluster 3. This figure indicates that frequency bands $A$, $B$, and $C$ are used by the edge zones of macrocells 1, 2, and 3, respectively. A small cell in a macrocell is not allowed to use the frequency band that is allocated to the overlaid macrocell, thus reducing cross-tier interference. Therefore, based on our frequency-allocation policy, the small cells in different macrocells can use one or more of the frequency bands $Z$, $A$, $B$, and $C$ apart from the overlaid frequency band or bands. The small cells in the center zones are not allowed to use the frequency band $Z$ as all the overlaid center zones of the macrocells use this frequency band. The allocated frequency bands for the small cells in the edge zones are ($Z$, $B$, $C$), ($Z$, $A$, $C$), and ($Z$, $A$, $B$) for macrocells 1, 2, and 3, respectively. Hence, $S_s^C \subset \{A, B, C\}$ is the set of allowed frequency bands for small cells in a center zone, and $S_s^E \subset \{Z, A, B, C\}$ is the set of allowed frequency bands for small cells in an edge zone. The allocated frequency band for each of the small cells in different macrocells can be expressed as

$$F_{s,1}^C \in \{B, C, B+C\} \tag{1}$$

$$F_{s,1}^E \in \{Z, B, C, Z+B+C, Z+B, Z+C, B+C\} \tag{2}$$

$$F_{s,2}^C \in \{A, C, A+C\} \tag{3}$$

$$F_{s,2}^E \in \{Z, A, C, Z+A+C, Z+B, Z+C, A+C\} \tag{4}$$

$$F_{s,3}^C \in \{A, B, A+B\} \tag{5}$$

$$F_{s,3}^E \in \{Z, A, B, Z+A+B, Z+A, Z+B, A+B\} \tag{6}$$

where $F_{s,i}^C$ and $F_{s,i}^E$ are the allocated frequency bands for a small cell situated in the center and edge zones, respectively, of the $i^{th}$ macrocell.

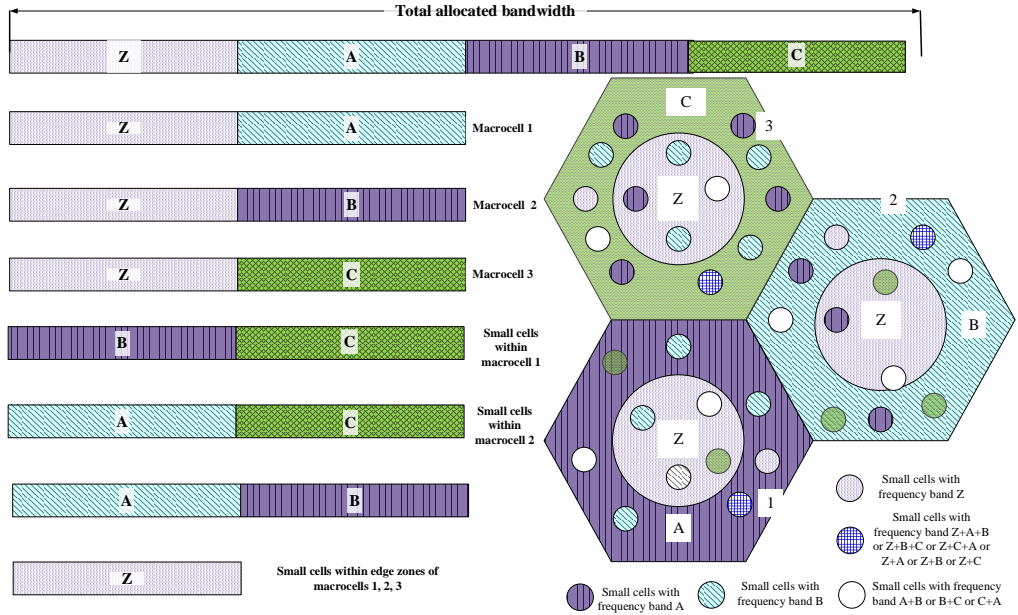

**Fig. 2.** Frequency-band allocation for macrocells and small cells in a three-macrocell cluster.

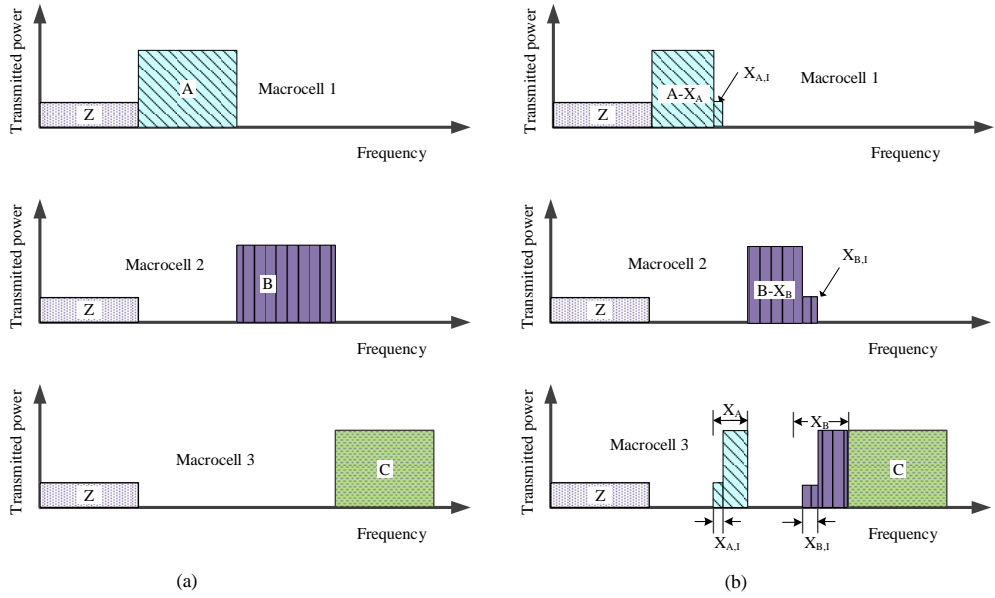

**Fig. 3.** Power allocation for three neighboring macrocells (a) before frequency-band reassignment and (b) after frequency-band reassignment.

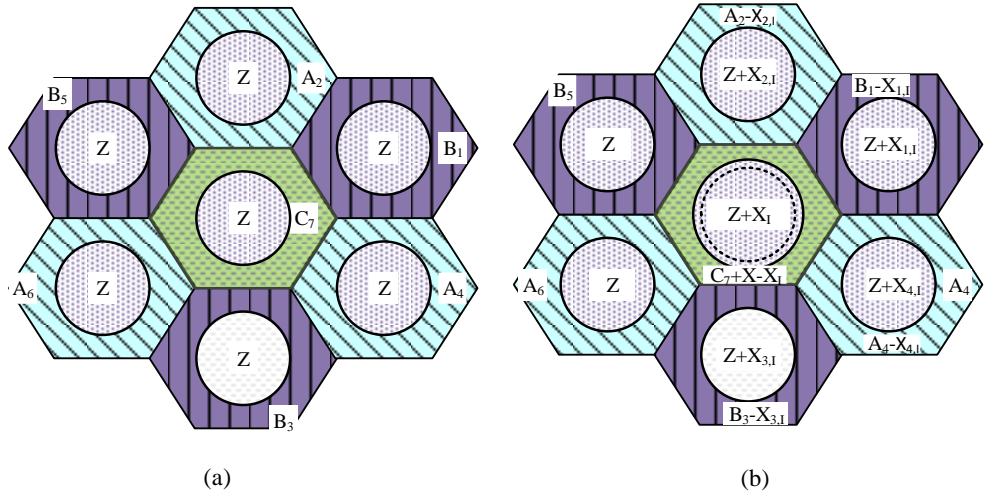

**Fig. 4.** Frequency-band allocation among cells of a seven-macrocell cluster (a) before frequency-band reassignment and (b) after frequency-band reassignment.

Transmitted power allocation can be used to significantly reduce interference. Fig. 3 shows the power allocation of three neighboring macrocells. The frequency bands that provide the connectivity for the edge zones are allocated maximum transmitted power $P_{t,M}$. All the macrocells use the $Z$ band to provide connectivity to their center-zone users. Therefore, the transmitted power $\alpha P_{t,M}$ for the $Z$ band is lower than $P_{t,M}$. Here, the coefficient $\alpha$ ($0 < \alpha < 1$) determines the transmitted power of the center zone and its coverage area compared to that of the edge zone and its coverage area. We propose that the frequency bands for different edge zones can be reassigned based on the traffic intensity in different macrocells instead of assigning fixed frequency bands $A$, $B$, and $C$ to macrocells 1, 2, and 3, respectively. For the reassignment of frequency bands, we assume that macrocell 3 has shortage of bandwidth to serve its users, whereas macrocell 1 and macrocell 2 have unused bandwidth. For the proposed frequency-allocation technique, a part of $A$ or $B$ or both frequency bands ($X_A$ of $A$ and $X_B$ of $B$) are allocated to serve the MUE in macrocell 3. Two cases can be occurred due to the frequency band reassigning to macrocell 3 (i) partial bands $X_A$ and $X_B$ in macrocells 1, and 2, respectively are totally unused, and (ii) a part of $X_A$ or $X_B$ or both is already occupied by existing users. Hence, these partial bands $X_A \in A$ and $X_B \in B$ in macrocell 3 may cause interference (above mentioned second case will cause interference) with macrocells 1 and 2, respectively. We assume that $X_{A,I} \in X_A$ and $X_{B,I} \in X_B$ are the frequency bands which interfere with macrocell 1 and macrocell 2, respectively. Hence, after frequency reassignment, the proposed scheme allocates power to different frequency bands in such a way that they cause minimum interference. The allocated powers of the interfering frequency bands $X_{A,I}$ and $X_{B,I}$ in different macrocells are kept low, as for the $Z$ band so that interference is minimized.

After reassigning the frequency bands, the number of users in the system is increased. However, improper frequency and power allocation may cause severe interference. Let us consider a cluster of seven macrocells with frequency reuse factors 1 and 3 for the center and edge zones, respectively. Fig. 4 shows the frequency allocation among different macrocells based on the FFR scheme. Fig. 4(a) shows that all seven macrocells use the $Z$ band for their center zone and one of the $A$, $B$, and $C$ bands for their edge zone. The following observations can be made: macrocell 7 uses frequency band $C$; macrocells 1, 3, and 5 use frequency band $B$; macrocells 2, 4, and 6 use frequency band $A$. Let us assume that macrocell 7 has a shortage of bandwidth to serve incoming users. We also assume that a part of frequency band $A$ ($X_A \in A$) in any of the macrocells 2, 4, and 6 and/or a part of frequency band $B$ ($X_B \in B$) in any of the macrocells 1, 3, and 5 are unused. Either of these bands $X_A$ and $X_B$ or both (i.e., $X = X_A + X_B$) can be reassigned to serve the MUE in macrocell 7. Fig. 4(b) shows the frequency-band allocation among seven macrocells after reassigning the frequency bands. We assume that macrocell 6 and macrocell 5 have the maximum unused bandwidth from $A$ and $B$, respectively. Therefore, these two macrocells will be unaffected by interference due to reassignment of frequency bands. In Fig. 4(b), $X_{i,I}$ (where $i \leq 4$) represents the frequency band of the $i^{th}$ macrocell that is affected by the reassignment. $X_I \in X$ is the frequency band that causes interference with the $X_{i,I}$ band of the $i^{th}$ macrocell (where $i \leq 4$). Therefore, all the interfering frequency bands $X_I$ and $X_{i,I}$ in the $7^{th}$ and $i^{th}$ macrocells, respectively, are transmitted to provide coverage for the center zones of these macrocells. We note that the transmitted power of the center-zone coverage can be readjusted if required.

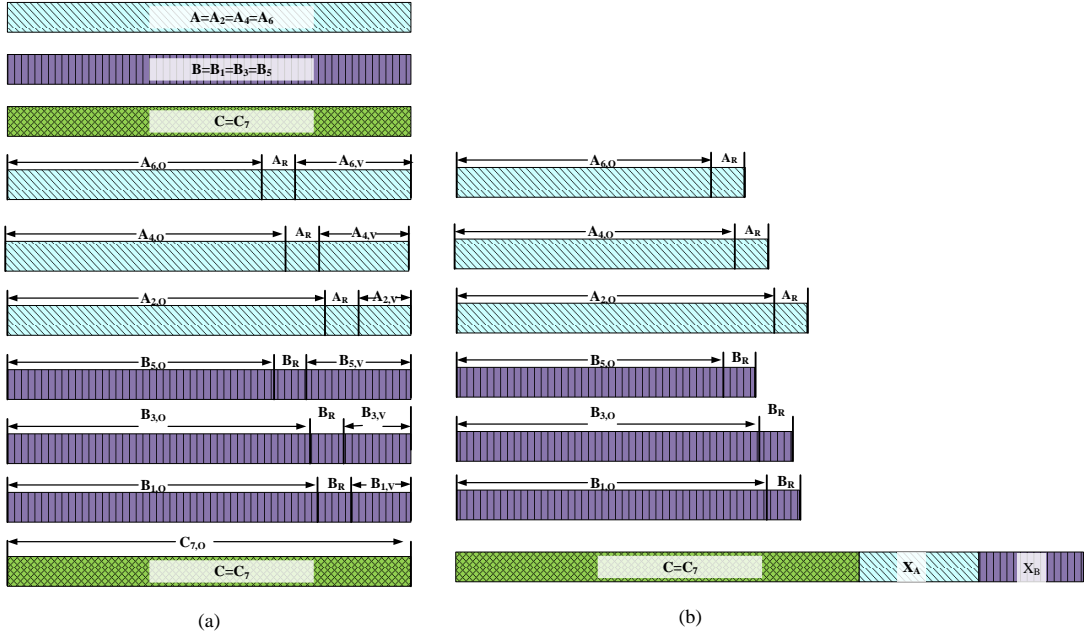

**Fig. 5.** Utility and reassignment conditions of frequency bands *A*, *B* and *C* (a) before frequency-band reassignment and (b) after frequency band reassignment.

SFR system is more bandwidth efficient compared to FFR system. It uses all the frequencies in each cell by using reuse factor [7]. However, interference level in FFR system is lower than SFR system. Moreover, SFR system is much more complex compared to FFR system. Even though, we apply FFR in our proposed scheme, the bandwidth utilization is maximized by allocating (i) dynamic frequency bandwidths among the macrocells based on their traffic load and (ii) remaining frequency bands to small cells that are not used by the overlaid macrocell. Hence, we consider currently utility conditions of re-used frequency bands for the proposed scheme. Fig. 5 shows the currently utility conditions of frequency bands *A*, *B*, and *C* and whether they can be reassigned to different macrocells. Before frequency-band reassignment, frequency band *A* is allocated to macrocells 2, 4, and 6, frequency band *B* is allocated to macrocells 1, 3, and 5, and frequency band *C* is allocated to macrocell 7. One or more of macrocells 1–6 may have some unused bandwidth. For $1 \leq i \leq 6$, $A_{i,O}$ (or $B_{i,O}$), $A_R$ (or $B_R$), and $A_{i,V}$ (or $B_{i,V}$) are the occupied, reserved, and releasable frequency bands in the $i^{th}$ macrocell, respectively. The reserve bandwidth is a part of unoccupied bandwidth that cannot be borrowed to others. This amount of bandwidth is kept backup to serve upcoming calls in that macrocell. The releasable frequency band is the remaining part of unoccupied bandwidth that can be borrowed to others. The addition of this releasable amount to other macrocell can increase the number of user connectivity. Macrocell 7 has a shortage of bandwidth for supporting incoming users. We assume that $A_{6,O} \leq A_{4,O} \leq A_{2,O}$, $B_{5,O} \leq B_{3,O} \leq B_{1,O}$, and $A_{6,O} \leq B_{5,O}$. Then, the maximum $A_{6,V} + B_{5,V}$ amount of bandwidth can be reallocated to macrocell 7 to support its incoming traffic calls.

Fig. 6 shows the frequency-band allocation among different macrocells and small cells with the macrocells after reassigning process. The frequency-bands are allocated in such a manner that the spectrum utilization can be increased significantly without increasing the interference effect.

It seems that owing to the reassignment of frequency bands there are multiple frequency-band options to allocate different traffic calls to in a macrocell. However, the link qualities of the frequency bands are not identical; some cause higher levels of interference compared to others. Therefore, there is scope to allocate the better-quality frequency bands to higher-priority calls. Fig. 7 shows a communication-link scenario for RT and nRT traffic calls with different frequency bands in the center and edge zones of macrocell 7. It shows that the reassigned frequency band $X$-$X_I$ is preferred for the RT traffic calls. However, the interfering frequency band $X_I$ is allocated only to nRT traffic calls in center zone. Thus, the proposed frequency-allocation scheme ensures a better QoS level for RT traffic calls. The interfering frequency band $X_I$ is also transmitted with lower power to serve the center-zone MUE. The center zone and edge zone small cells use frequency bands (*A*, *B*) and (*Z*, *A*, *B*), respectively, before channel reassignment. On the other hand, frequency bands ($A$-$X_{A,I}$, $B$-$X_{B,I}$) and ($Z$, $A$-$X_A$+$X_{A,I}$, $B$- $X_B$+$X_{B,I}$), respectively, are used for the center zone and edge zone small cells after channel reassignment. Thus, the proposed scheme avoids cross-tier interferences.

The bandwidth allocation is performed by cooperating among MBSs and sBSs. It is a well-known established process and there are many existing bandwidth allocation schemes (e.g., [31], [33]-[36]) in wireless communications. Therefore, the existing resource allocation procedure is also applicable for the bandwidth allocation of our proposed scheme. We provide a generalized frequency allocation scheme for OFDMA based cellular networks. Hence, it is applicable for any OFDMA based cellular network.

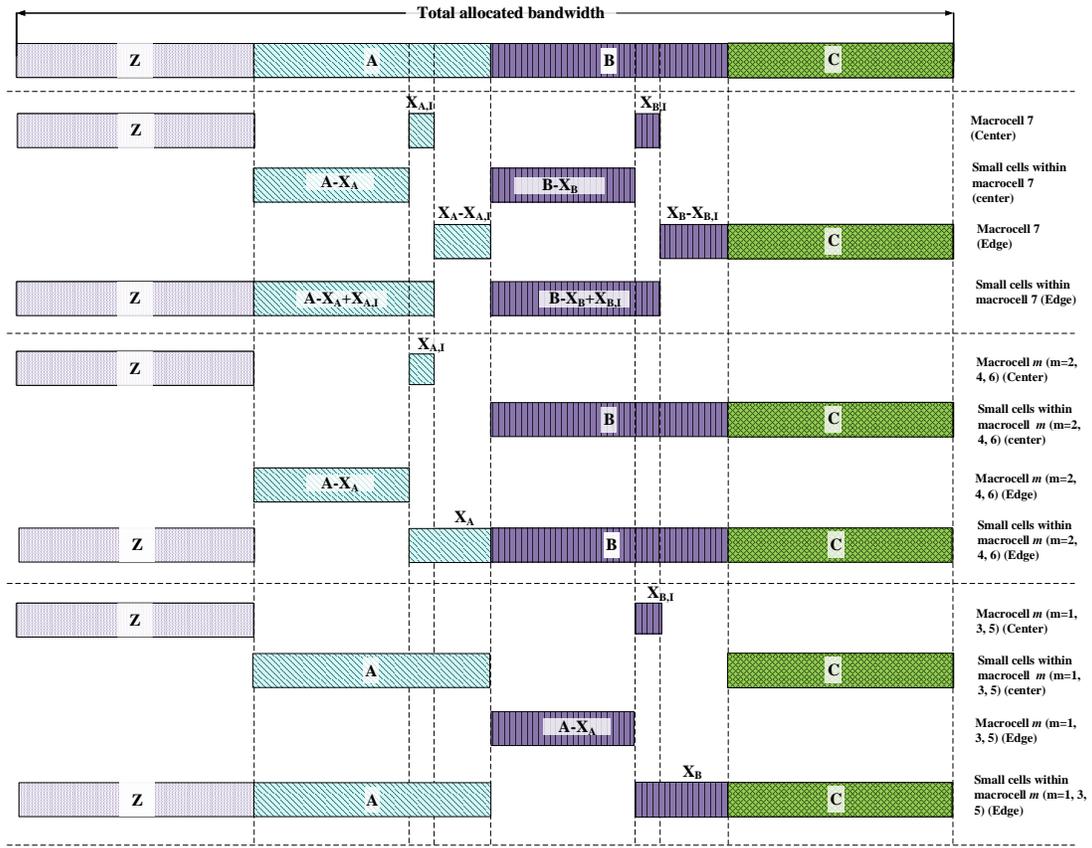

**Fig. 6.** Frequency-band allocation among different macrocells and small cells with the macrocells after reassigning process.

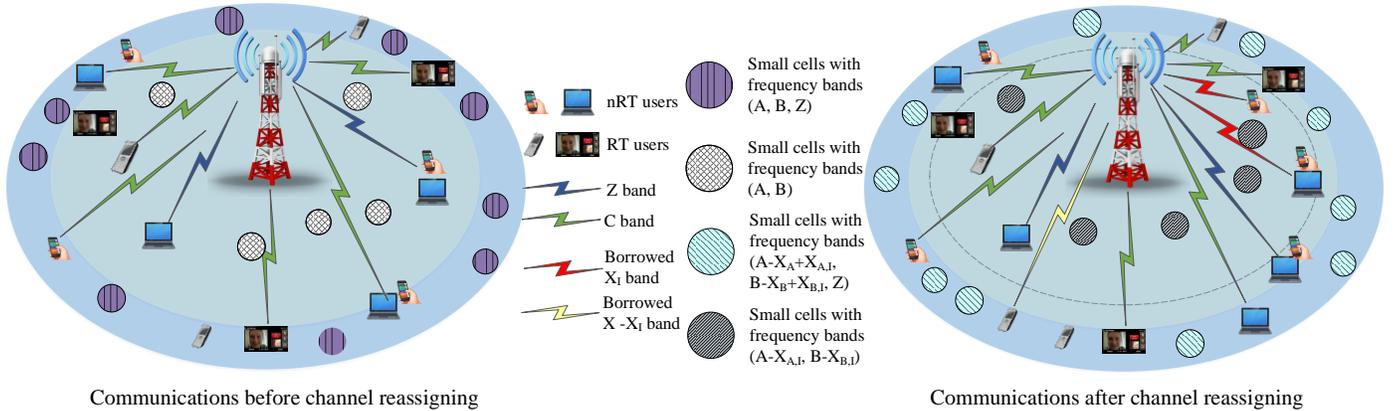

**Fig. 7.** Communication-link scenario for RT and nRT traffic with different frequency bands in the center and edge zones of a macrocell cluster before and after frequency-band reassignment

Our proposed frequency-allocation technique considers traffic classification within the FFR technique. The following two situations can occur: either the assigned frequency bands are sufficient to serve the incoming calls or they are insufficient. Therefore, the frequency-allocation process is different in these two scenarios. Fig. 8 shows the assignment procedure between RT and nRT users when the frequency bandwidth is sufficient (i.e., Z and $C_7$ frequency bands are sufficient to serve the users of macrocell 7). There is no frequency reassignment in this case. The available frequency bands Z and $C_7$ (= C) are allocated based on the traffic type. In the edge zone, only frequency band C is available. However, in the center zone, both frequency bands are available. The allocation of frequency band C causes less interference compared to the Z band. Hence, all RT and nRT traffic calls within the edge zone are allocated to frequency band C. The center zone traffic calls are then supported using the remainder of band C (if it exists) and band Z. The remainder of frequency band C is assigned preferentially to RT traffic calls. Hence, $BW_{req}$, $BW_{req,RT}$, $BW_{req,nRT(E)}$, and $BW_{req,nRT}$ are the required bandwidths for all calls, RT calls, nRT calls in the edge zone, and nRT calls, respectively.

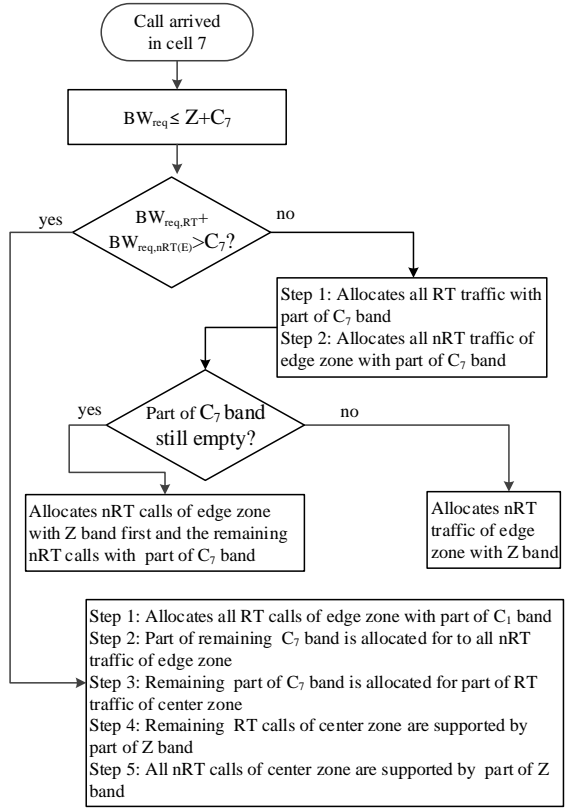

**Fig. 8.** Frequency-band assignment procedure between RT and nRT users when the required frequency bandwidth for cell 7 is not more than $Z + C_7$.

Fig. 9 shows the procedure for frequency-band assignment between RT and nRT users when the frequency bandwidth is not sufficient and frequency-band reassignment is required (i.e., $Z + C_7$ frequency bands are not sufficient to serve the users of macrocell 7). Frequency reassignment is performed according to demands. Then, the preassigned frequency bands $Z$ and $C_7$ and newly assigned frequency band $X$ are allocated based on the priority of traffic calls. Our scheme provides higher priority for RT calls than nRT traffic calls. The preassigned frequency band $C$ and reassigned $X-X_I$ band are available to edge zone users. However, in the center zone, all frequency bands ($C$, $Z$, $X-X_I$, and $X_I$) are available. The allocation of frequency bands $C$ and $X-X_I$ causes less interference compared to assigning frequency bands $Z$ and $X_I$. Therefore, the frequency bands $C$ and $X-X_I$ are preferred for RT traffic calls in both the center and edge zones. Center zone nRT traffic calls can access $C$ or $X-X_I$ frequency bands only when there is a remainder after assigning these bands to all edge zone users and the RT users in the center zone. Otherwise, they use the $Z$ or $X_I$ bands. We also note that if RT traffic calls require $Z$ or $X_I$ bands owing to lagging of the $C$ or $X-X_I$ frequency bands, then RT traffic calls will get preference over nRT traffic calls. Thus, our proposed scheme ensures a low call-blocking rate and high SINR for RT traffic calls.

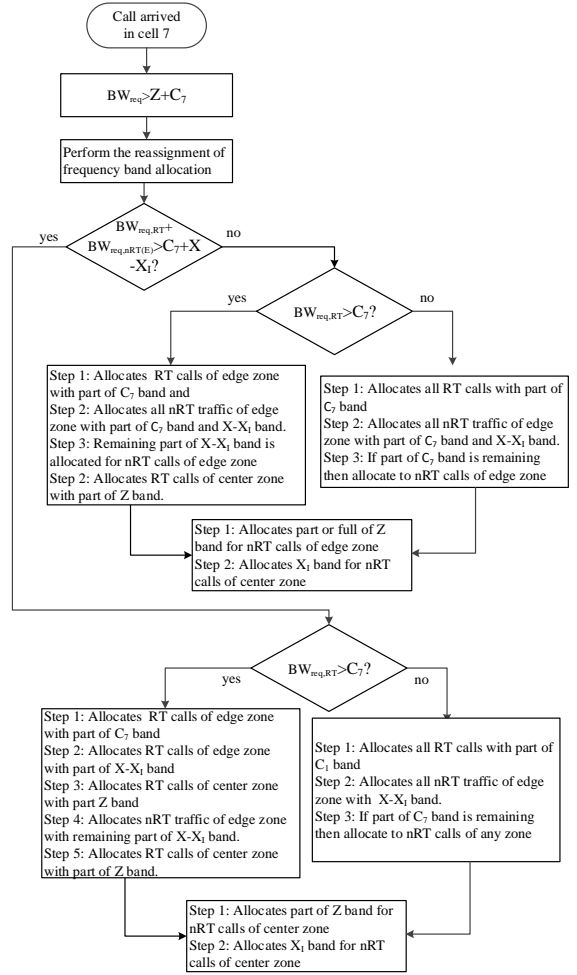

**Fig. 9.** Frequency-band assignment procedure between RT and nRT users when the required frequency bandwidth for cell 7 is more than $Z+C_7$.

## IV. INVESTIGATION OF INTERFERENCE EFFECT

The main objective of reassigning parts of bands $A$ and $B$ to macrocell 7 is to reduce the call-blocking rate as well as to increase the resource utilization. Owing to the reassignment of frequency bands among different macrocells, the reassigned bands may create interference between the macrocells. The reassignment can also cause interference for small cells. However, interference effects on or from small cells can easily be mitigated if the available frequency band for small cells is sufficiently large. The novel RT/nRT traffic classification based FFR seems to be more effective approach for macrocell and we are mainly focusing the analysis on macrocells in the HetNets. The interference due to the deployment of small cells can be mitigated using the conventional schemes (e.g., [1], [32], [37]-[39]). We also have several research works (e.g, [5], [32]) on interference management for small cell network deployment that can be applied for this interference mitigation. Fig. 10 shows some examples of the affected frequency sub-bands of various macrocells due to reassigning frequency band $A$. Interference effects due to reassigning frequency band $B$ can be explained in the same manner. The size of the affected frequency band in a macrocell depends on the amount

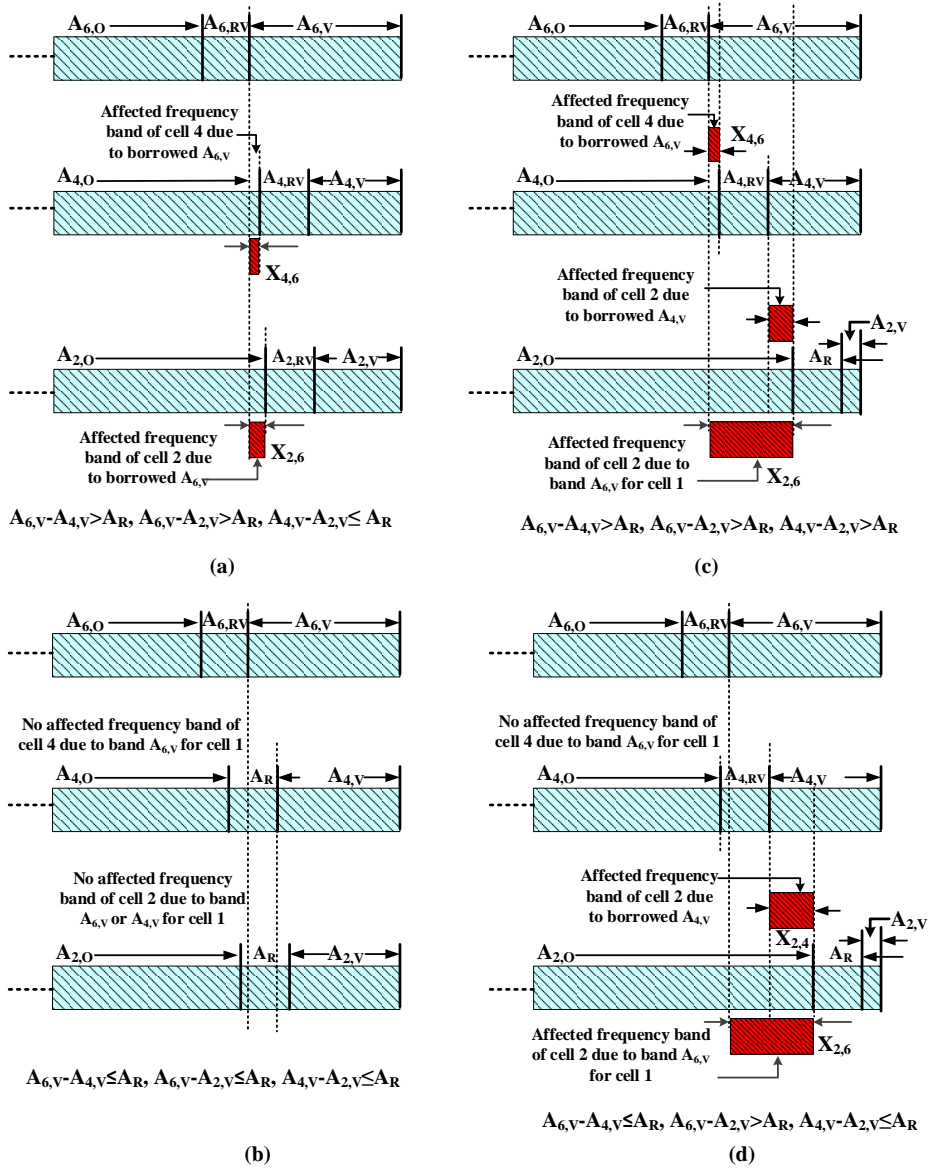

**Fig. 10.** Examples of the affected frequency sub-bands of various macrocells due to reassigning frequency band *A*.

of reassigned bandwidth, the amount of bandwidth already being used in that macrocell, and the amount of bandwidth reserved for future incoming traffic. From Fig. 10, it seems that interference is not always caused by the reassigned frequency bands. In addition, if the amount of occupied bandwidth in a macrocell is high, the frequency band is greatly affected by the reassignment process. The affected frequency bands from different macrocells should be handled carefully to minimize interference. In our proposed scheme, we transmit these interference-causing frequency bands with low power and assign them to low-priority nRT traffic calls, which have more flexibility in their QoS requirements

Fig. 11 shows case studies of interference in various macrocells due to reassigning frequency bands *A* and *B*. The figure clearly shows that the magnitude of the interference effects due to frequency-band reassignment depends on several criteria. No interference effects are observed in the other macrocells when any of the frequency bands $A_{2,V}$, $B_{1,V}$, or $A_{2,V} + B_{1,V}$ is reassigned to macrocell 7. However, the maximum number of macrocells (macrocells 1, 2, 3, and 4) is affected when the $A_{6,V} + B_{5,V}$ frequency band is reassigned to macrocell 7. Therefore, based on the assigned and interfering frequency bands, the non-interfering and interfering frequency bands can be allocated to RT and nRT traffic calls, respectively, to ensure a better QoS level for RT traffic calls.

Our proposed interference management scheme considers frequency availability, availability for frequency reassignment, zones of the macrocell, and traffic type. Fig. 12 shows the basic procedure for the proposed scheme that is based on FFR and RT/nRT traffic classification. Fig. 12(a) shows the five consequence steps for frequency allocation. We also provide the grouping of frequency bands according to preference for traffic class so we can ensure a better QoS level in terms of low call- blocking rate, higher SINR, and lower connection outage

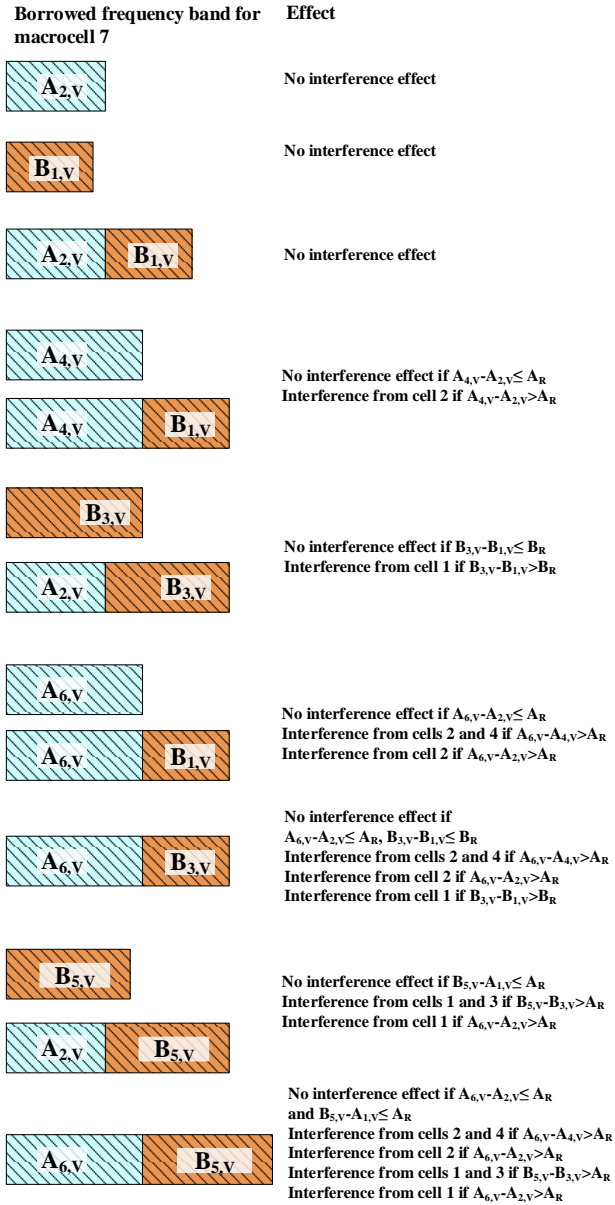

**Fig. 11.** Case studies of interference in various macrocells due to reassigning frequency bands *A* and *B*.

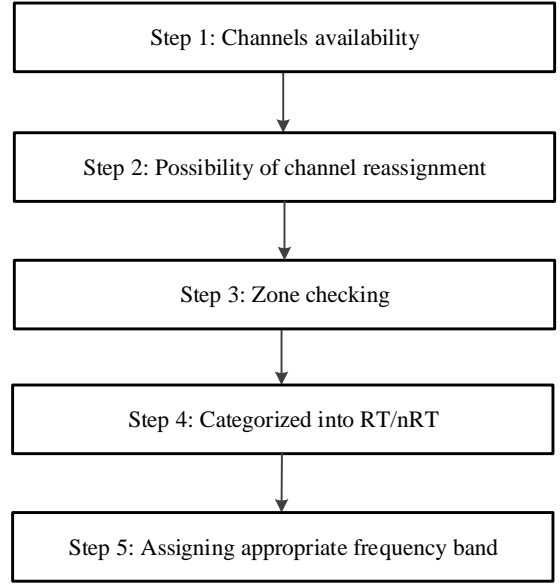

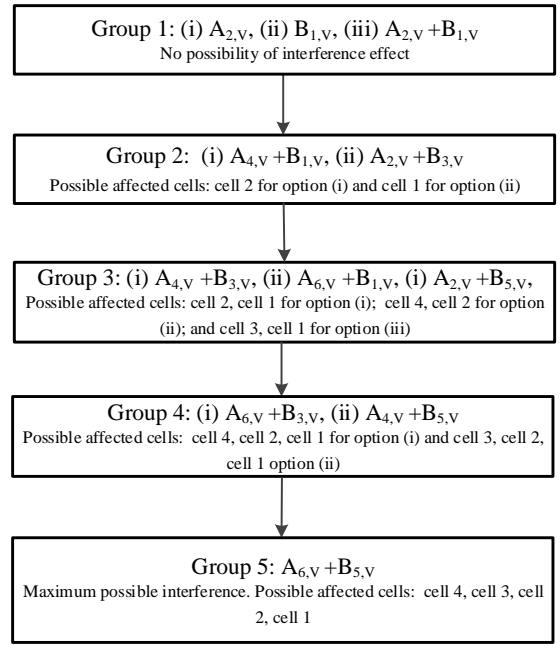

**Fig. 12.** Basic procedure for the proposed frequency allocation (a) five consequence steps and (b) grouping of frequency sub-bands according to preference.

probability for higher-priority calls. Different frequency bands cause different level of interferences. Therefore, we categorized different frequency bands into five groups. In terms of interference, the effect of first group is lowest and which is gradually increased up to fifth group. Hence, group 1 frequency bands are most preferable for frequency reassignment. Fig. 12(b) shows five different groups of frequency bands for macrocell 7, according to their preference for reassignment. The system reassigns the frequency bands from the group 1 first. If more frequency is demanded, then it moves to next group. The five steps of frequency-band allocation are briefly described as follows:

Step 1: The system checks whether channels are available to support the requested call. If channels are available, the system accepts the call. Otherwise it moves to step 2.

Step 2: The system verifies whether channel reassignment is possible. If it is not possible, the system rejects the call request and terminates the process. If channel reassignment is possible, it borrows channels from band *A*, band *B* or both and goes to step 3.

Step 3: The system confirms whether the user is in the center or edge zone. Channel allocation for center and edge zones is different, and available channels are also different in the two zones.

Step 4: The system checks whether the type of call is RT or nRT. The QoS requirements for the two are different. Therefore,

RT calls will be preferentially allocated to lower-interference frequency bands. After checking the traffic type, the system moves to step 5 for frequency-band allocation.

Step 5: The system checks all the available frequency bands that can support the requested call. Based on the available frequency bands, traffic type, and priority of frequency band, the system assigns the appropriate frequency band to the call.

Owing to the reassignment of frequency bands, the criteria for the assigned frequency bands for macrocell 7 can be written as

$$|F_7| = |Z| + |C| + |X| \quad (7)$$

$$|X| = |X_A \in A| + |X_B \in B| \quad (8)$$

$$|F_7^C| = |Z| + |X_I| \quad (9)$$

$$|F_7^E| = |C| + |X| - |X_I| \quad (10)$$

$$|X_I| = |X_A \cap \{\min(X_2, X_4, X_6)\}| + |X_B \cap \{\min(X_1, X_3, X_5)\}| \quad (11)$$

$$\sum_{j=1}^{6} F_7^E \cap F_j^E = \phi \quad (12)$$

$$\sum_{j=1}^{6} F_7^E \cap F_j^C = \phi \quad (13)$$

where $F_i^C$ and $F_i^E$ are the total allocated frequency bands for the center and edge zones, respectively, of the $i^{th}$ macrocell. $X_{i,I}$ is the frequency sub-band of the $i^{th}$ macrocell that may interfere with macrocell 7. The suffix $j$ indicates the $j^{th}$ neighbor MBS from which the MUE receives interference power.

The received SINR level for the $k^{th}$ MUE in the $i^{th}$ macrocell can be expressed as (14)-(18). In (14)-(18), $SINR_{k,i(C)}$ and $SINR_{k,i(E)}$ are the received SINR level for the $k^{th}$ MUE in the center and edge zones, respectively, of the $i^{th}$ macrocell, $h_{k,i(C)}$ and $h_{k,i(E)}$ are the channel gains between the $k^{th}$ MUE and the center and edge zone antennas of the MBS, respectively, $P_{t,s}$ is the power transmitted by an sBS, and $h_{k,s}$ is the channel gain between the MUE and sBS. The suffixes $j$ and $m$ indicate the $j^{th}$ neighbor MBS and $m^{th}$ neighbor sBS, respectively, from which the MUE receives interference power. $N_0$ is the received noise power to an MUE.

The received SINR level for the $k^{th}$ RT and nRT MUE in macrocell 7 can be expressed as (19)-(22) where $I_{2+}$ is the received interference power from $2^{nd}$- and higher-tier macrocells.

$$SINR_{k,i(C)} = \frac{\alpha P_{t,M} |h_{k,i(C)}|^2}{\sum_{j \neq i} \delta_{k,j(C)} |h_{k,j(C)}|^2 \alpha P_{t,M} + \sum_{j \neq i} \delta_{k,j(E)} |h_{k,j(E)}|^2 P_{t,M} + \sum_m \delta_{k,m,s} |h_{k,s}|^2 P_{t,s} + N_0} \quad (14)$$

$$SINR_{k,i(E)} = \frac{P_{t,M} |h_{k,i,(E)}|^2}{\sum_{j \neq i} \delta_{k,j(C)} |h_{k,j(C)}|^2 \alpha P_{t,M} + \sum_{j \neq i} \delta_{k,j(E)} |h_{k,j(E)}|^2 P_{t,M} + \sum_m \delta_{k,m,s} |h_{k,s}|^2 P_{t,s} + N_0} \quad (15)$$

$$\delta_{k,j,(C)} = \begin{cases} 1 & \text{if } k^{th} \text{ MUE in center zone is associated with center zone} \\ & \text{of } j^{th} \text{ MBS with the same frequency} \\ 0 & \text{otherwise.} \end{cases} \quad (16)$$

$$\delta_{k,j,(E)} = \begin{cases} 1 & \text{if } k^{th} \text{ MUE in edge zone is associated with edge zone} \\ & \text{of } j^{th} \text{ MBS with the same frequency} \\ 0 & \text{otherwise.} \end{cases} \quad (17)$$

$$\delta_{k,m,s} = \begin{cases} 1 & \text{if } k^{th} \text{ MUE is associated with } m^{th} \text{ sBS with} \\ & \text{the same frequency} \\ 0 & \text{otherwise.} \end{cases} \quad (18)$$

$$SINR_{k,7(C)}^{nRT} = \frac{\alpha P_{t,M} |h_{k,7,(C)}|^2}{\sum_{j:1 \leq j \leq 6} |h_{k,j(C)}|^2 \alpha P_{t,M} + \sum_{j:1 \leq j \leq 6, j \notin \{5,6\}} |h_{k,j(C)}|^2 P_{t,M} + \sum_m \delta_{k,m,s} |h_{k,s}|^2 P_{t,s} + I_{2+} + N_0} \quad (19)$$

$$SINR^{RT}_{k,7(C)} = \frac{\alpha P_{t,M} |h_{k,7,(C)}|^2}{\sum_{j:1\leq j\leq 6} |h_{k,j(C)}|^2 \alpha P_{t,M} + I_{2+} + N_0} \quad (20)$$

$$SINR^{nRT}_{k,7(E)} = \frac{P_{t,M} |h_{k,7,(E)}|^2}{\sum_{j:1\leq j\leq 6, j\notin\{5,6\}} \delta_{k,j(E)} |h_{k,j(E)}|^2 P_{t,M} + \sum_m \delta_{k,m,s} |h_{k,s}|^2 P_{t,s} + I_{2+} + N_0} \quad (21)$$

$$SINR^{RT}_{k,7(E)} = \frac{P_{t,M} |h_{k,7,(E)}|^2}{I_{2+} + N_0} \quad (22)$$

The achievable data rates $R_{k,i(C)}$ and $R_{k,i(E)}$ for the $k^{th}$ MUE in the center and edge zones, respectively, of the $i^{th}$ macrocell in [bits/sec/Hz] are given as

$$R_{k,i(C)} = \log(1 + SINR_{k,i(C)}) \quad (23)$$

$$R_{k,i(E)} = \log(1 + SINR_{k,i(E)}) \quad (24)$$

If $N_C$ and $N_E$ are the number of channels allocated to the center and edge zones of a macrocell, respectively, then the net area spectral efficiency in [bits/sec/Hz/macrocell] of that macrocell is given as

$$ASE_{macro} = \frac{N_C}{N_C + N_E} \frac{C_C}{\Delta_C} + \frac{N_E}{N_C + N_E} \frac{C_E}{\Delta_E} \quad (25)$$

where $\Delta_C$ and $\Delta_E$ are the frequency reuse factors for the center and edge zones, respectively. $C_C$ and $C_E$ are the capacity per unit frequency in bits/sec/Hz for the center and edge zones, respectively.

Due to the relocation of frequency bands, the criteria for the assigned frequency bands for small cells within macrocell 7 can be written as

$$|F_{7(s)}| = |Z| + |A| + |B| \quad (26)$$

$$|F_{7(s,C)}| = |A - X_A| + |B - X_B| \quad (27)$$

$$|F_{7(s,E)}| = |Z| + |A - X_A| + |B - X_B| + |X_I| \quad (28)$$

$$F_7^C \cap F_{7(s,C)} = \phi \quad (29)$$

$$F_7^E \cap F_{7(s,E)} = \phi \quad (30)$$

where $F_{7(s)}$, $F_{7(s,C)}$, and $F_{7(s,E)}$ are the total allocated frequency bands for the small cells within macrocell 7, center zone of macrocell 7, and edge zone of macrocell 7, respectively.

The received SINR level for the $n^{th}$ sUE in the $i^{th}$ macrocell can be expressed as (31) and (32) where $SINR_{n,i(C)}$ and $SINR_{n,i(E)}$ are the received SINR level for the $n^{th}$ sUE in the small cell within center and edge zones, respectively, of the $i^{th}$ macrocell; $h_{n,i(C)}$ and $h_{n,i(E)}$ are the channel gains between the $n^{th}$ sUE and the center and edge zone antennas of the MBS, respectively. The received SINR level for the $n^{th}$ sUE in macrocell 7 can be expressed as (33) and (34)

$$SINR_{n,i(C)} = \frac{P_{t,s} |h_{n,s}|^2}{\sum_{j\neq i} \delta_{n,j(E)} |h_{n,j(E)}|^2 P_{t,M} + \sum_m \delta_{n,(m,s)} |h_{k,s}|^2 P_{t,s} + N_0} \quad (31)$$

$$SINR_{n,i(E)} = \frac{P_{t,s} |h_{n,s}|^2}{\sum_{j\neq i} \delta_{n,j(C)} |h_{n,j(C)}|^2 \alpha P_{t,M} + \sum_{j\neq i} \delta_{n,j(E)} |h_{n,j(E)}|^2 P_{t,M} + \sum_m \delta_{n,(m,s)} |h_{k,s}|^2 P_{t,s} + N_0} \quad (32)$$

$$SINR_{n,i(C)} = \frac{P_{t,s} |h_{n,s}|^2}{\sum_{j:1\leq j\leq 6} \delta_{n,j(E)} |h_{n,j(E)}|^2 P_{t,M} + \sum_m \delta_{n,(m,s)} |h_{k,s}|^2 P_{t,s} + N_0} \quad (33)$$

$$SINR_{n,7(E)} = \frac{P_{t,s} |h_{n,s}|^2}{\sum_{j:1\leq j\leq 6} \delta_{n,j(C)} |h_{n,j(C)}|^2 \alpha P_{t,M} + \sum_{j:1\leq j\leq 6} \delta_{n,j(E)} |h_{n,j(E)}|^2 P_{t,M} + \sum_m \delta_{n,(m,s)} |h_{k,s}|^2 P_{t,s} + I_{2+} + N_0} \quad (34)$$

where the value of each of the terms $\delta_{n,j(C)}$, $\delta_{n,j(C)}$, and $\delta_{n,j(C)}$ is either 1 or 0. The conditions $\delta_{n,j(C)} = 1$, $\delta_{n,j(E)} = 1$, and $\delta_{n,(m,s)} = 1$, respectively, satisfy only if $n^{th}$ sUE associated with center zone of $j^{th}$ MBS, edge zone of $j^{th}$ MBS, and $m^{th}$ neighbor sBS with the same frequency.

The Okumura-Hata path loss model [40] can be used to ascertain path losses in macrocells. The path loss for a macrocell can be expressed as

$$L_{macro} = 69.55 + 26.16 \log f_c - 13.82 \log h_b - a(h_m) + (44.9 - 6.55 \log h_b) \log d + L_{ow} \quad [dB] \quad (35)$$

$$a(h_m) = 1.1(\log f_c - 0.7)h_m - (1.56 \log f_c - 0.8) \quad [dB] \quad (36)$$

where $d$ is the distance between the MBS and the user in kilometers, $f_c$ is the center frequency of the antenna in MHz, $h_m$ is the height of user in meters, $h_b$ is the height of MBS in meters, and $L_{ow}$ represents the penetration loss.

The path loss for a small cell can be expressed as [41]

$$L_{femto} = 20 \log f + N \log z + L_f(n) - 28 \quad [dB] \quad (37)$$

where $z$ is the distance between the sBS and the user in meters, $f$ is the center frequency of the antenna in MHz, $N$ is the distance power loss coefficient, and $L_f$ is the floor penetration loss factor.

The connection outage probability of the $k^{th}$ MUE in the $i^{th}$ macrocell can be expressed as

$$P_{outage,k,i} = \Pr\left(SINR_{k,i} < \gamma\right) \quad (38)$$

where $\gamma$ is the threshold value of the SINR.

Equation (29) can then be written as

$$P_{outage,k,i} = 1 - \exp\left[-\frac{\gamma}{S_o}\left(\sum_{j \neq i} \delta_{k,j(C)} |h_{k,j}|^2 \alpha P_{t,M}\right.\right.$$
$$+ \sum_{j \neq i} \delta_{k,j(E)} |h_{k,j}|^2 P_{t,M} \quad (39)$$
$$\left.\left.+ \sum_m \delta_{k,m,s} |h_{k,s}|^2 P_{t,s} + N_0\right)\right]$$

where $S_0$ is the signal power received by an MUE from the MBS.

The proposed RT/nRT traffic classification for the FFR is a new concept and applicable in any situation (weather lagging of bandwidth or not) to assign frequency bands for RT and nRT traffic calls and also for center zone users and edge zone users. Hence, QoS for RT users is guaranteed in our scheme. However, the bandwidth reassigning process is done only when there is a shortage of bandwidth in any macrocell. Even though, we have shown few macrocells with empty bandwidth, however, we can also apply this scheme whenever all the cells in a cluster have lagging of bandwidth. However, for that case, the reassigned additional bandwidths should be assigned to center-zone nRT users.

TABLE II
PARAMETER VALUES USED IN THE PERFORMANCE ANALYSIS

| Parameter | Value |
|---|---|
| Center frequency | 1800 MHz |
| Transmitted signal power by the MBS for edge zone | 46 dBm |
| Transmitted signal power by the MBS for center zone | 35 dBm |
| Transmitted signal power by the sBS | 7 dBm |
| Noise power density | -174 dBm/Hz |
| Distance between two MBSs | 1000 m |
| Macrocell center zone radius | 250 m |
| Wall penetration loss | 10 dB |
| Threshold value of SINR ($\gamma$) | 8.45 dB |
| Height of the MBS | 50 m |
| Height of the UE | 2 m |
| Number of sBS deployed in each macrocell | 100 |
| Distance power loss coefficient (N) for small cells | 28 |

## V. PERFORMANCE ANALYSIS

In this section, we evaluate the performance of our proposed scheme by comparing our scheme with conventional FFR schemes in terms of network capacity, outage probability, call-blocking probability, and achievable SINR. The performance measurement parameters of the proposed scheme are compared to the conventional FFR approach, where frequency-band reassigning and traffic classification are not performed [6]-[11], [15], [16]. The results show the features and improvements provided by the proposed scheme, which ensures higher frequency utilization and better QoS level for higher-priority RT traffic calls. Interference effects from the 3$^{rd}$ tier and above are negligible. Therefore, we consider only the macrocells in the 1$^{st}$ and 2$^{nd}$ tiers of the reference macrocell and small cells within a 100 m range of the reference MUE. Table II summarizes the fundamental parameters that we used to define performance in our numerical analysis. We provide separate performance analyses for the center and edge zones since the power allocation for these two zones is different. We also show the performance of our proposed scheme for the RT/nRT traffic classification and no traffic classification cases. Macrocell 7 is the reference macrocell for the performance analysis.

Due to the reassignment of frequency bands, our proposed scheme surely increases the number of call connections. Our scheme improves the SINR performances as well. This is because, after reassigning the frequency bands, we properly allocate power to lower- and higher-interference frequency bands. Hence, our proposed scheme can outperform the conventional FFR scheme. It is possible to allocate all the frequency bands $C$, $Z$, $X-X_I$, and $X_I$ to the center-zone users; therefore, several band options are available for assigning to RT and nRT traffic calls. Among these frequency bands, some cause less interference than other bands. The proposed scheme preferentially assigns less-interfering frequency bands to RT traffic calls. As a result, RT traffic calls suffer from less interference and achieve better SINR levels compared to nRT traffic calls. The available frequency bands in the proposed scheme for MUE in the edge zone are $C$ and $X$-$X_I$ of macrocell 7. Our scheme also causes less interference to edge-zone MUE due to the proper allocation of power allocation. Consequently, we can achieve better capacity. Because of reassigning of frequency bands, a serious interference can be happened in conventional FFR. Fig. 13 illustrates a comparison of macrocell downlink capacities for different MUE in the center zone. The capacity of the proposed scheme is larger than that of the conventional FFR scheme. By adding RT/nRT traffic classification, the proposed scheme improves performance for RT traffic calls without sacrificing the performance of nRT traffic calls. The capacity for RT traffic calls increases significantly, which also ensures better QoS, and the capacity for nRT traffic calls also increases noticeably. Fig. 14 shows a comparison of macrocell downlink capacities for different MUE in the edge zone. The proposed scheme improves the capacity significantly. For the edge zone case, the available frequency bands $C$ and $X$-$X_I$ show similar interference characteristics. Therefore, the performance for the RT, nRT, and unclassified traffic cases are almost the same.

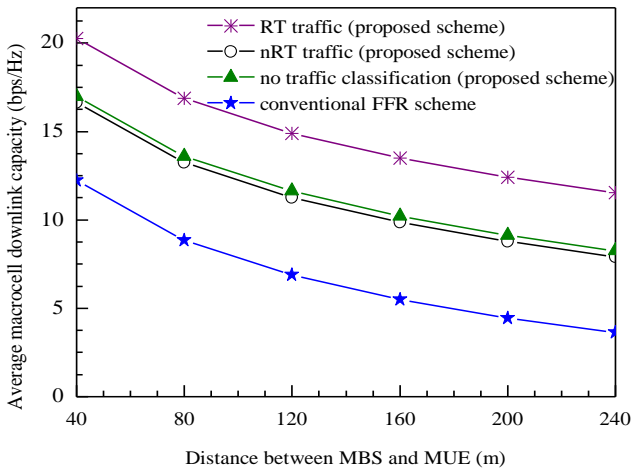

**Fig. 13.** Comparison of macrocell downlink capacities for different MUE in the center zone.

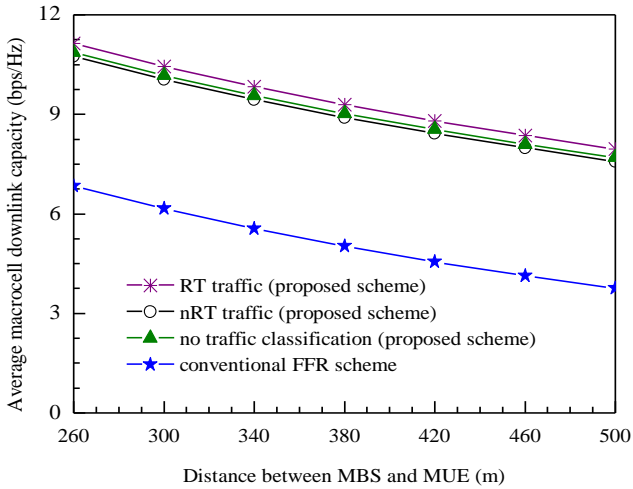

**Fig. 14.** Comparison of macrocell downlink capacities for different MUE in the edge zone.

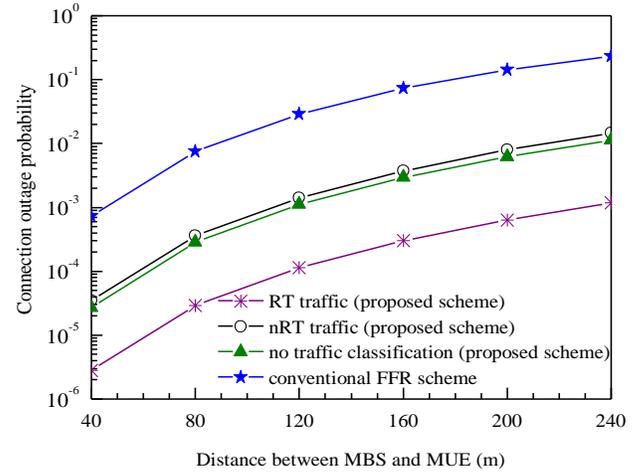

**Fig. 15.** Comparison of connection outage probabilities for different MUE in the center zone.

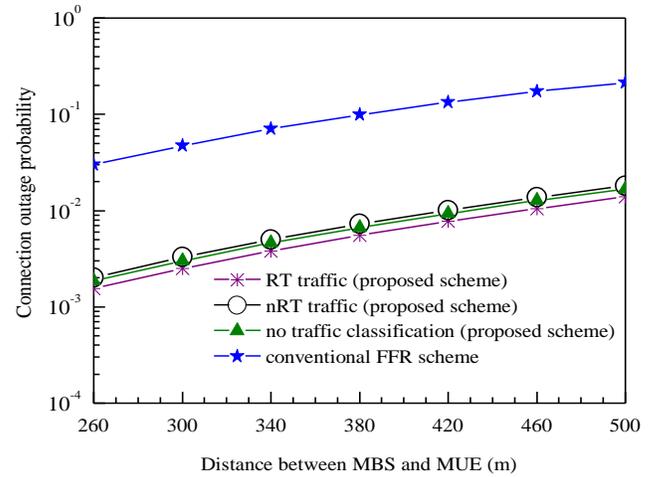

**Fig. 16.** Comparison of connection outage probabilities for different MUE in the edge zone.

The connection outage probability is defined as the probability that the SINR goes down the threshold level. It is an important metric to evaluate communication reliability. Figs. 15 and 16 show that our proposed interference management scheme also maintains the connection outage probability within the considered range. For the different MUE, the proposed scheme provides almost negligible connection outage probability for RT traffic calls and that of the nRT MUE is also within the acceptable range. Conventional FFR schemes cannot provide much smaller connection outage probabilities than the proposed scheme. For the edge zone case, the performance for the RT, nRT, and unclassified traffic under the proposed scheme is almost the same since the available frequency bands are provided with the same power and cause similar levels of interference in the MUE. However, the higher priority of RT traffic calls in frequency allocation reduces their call-blocking probability compared to nRT traffic calls.

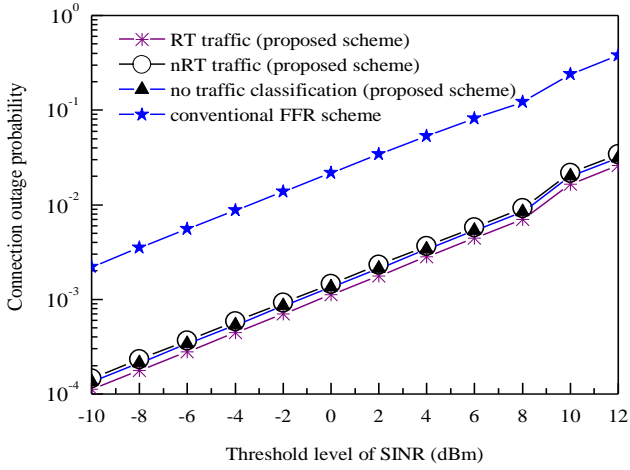

**Fig. 17.** Outage probability comparison for an MUE in the center zone (240 m from the MBS).

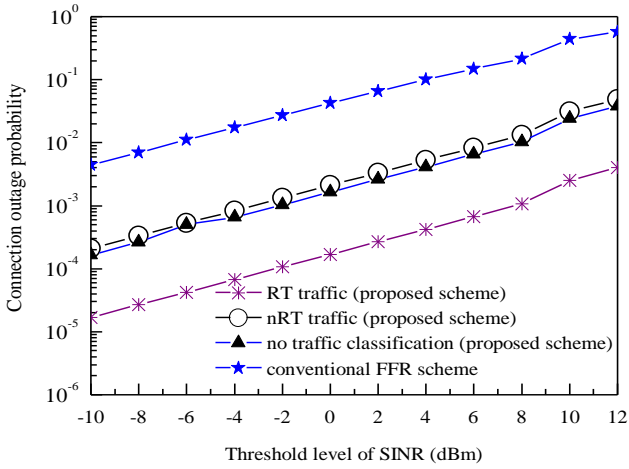

**Fig. 18.** Outage probability comparison for an MUE in the edge zone (420 m from the MBS).

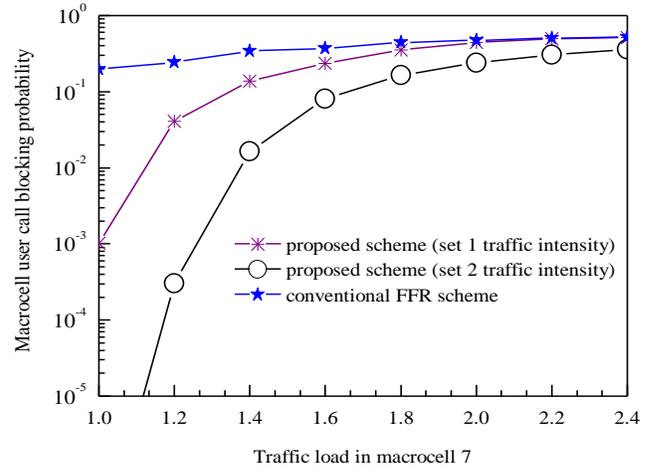

**Fig. 19.** Comparison of macrocell user call-blocking probability.

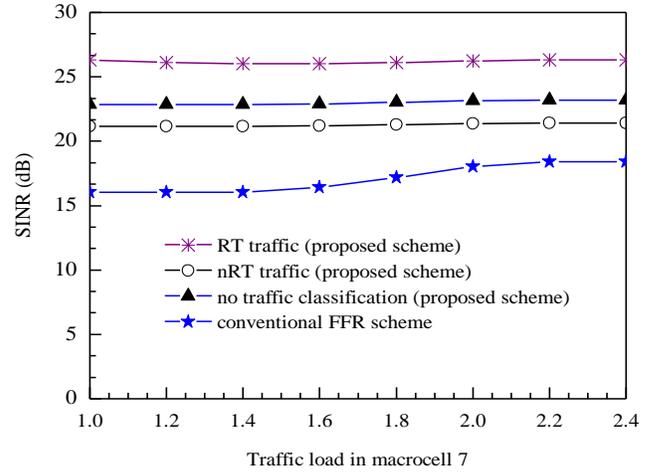

**Fig. 20.** Comparison of SINR for a MUE at center zone (240 m distance).

Figs. 17 and 18 show that the proposed scheme can perform better in terms of connection outage probability with the variation of SINR threshold level. Even though the threshold value of the SINR is increased to 12 dB, the connection outage probability for our scheme is still very low. However, conventional FFR schemes cannot maintain low connection outage probability when the SINR threshold is increased. We consider an MUE 240 m away from the MBS in the center zone case, which is near the boundary of the center zone. In Fig.17, we observe that the connection outage probability increases when the SINR threshold is increased. Our proposed scheme provides multiple options for frequency-band selection and transmits at a lower power for highly interfering frequency bands. Therefore, our proposed scheme provides a better SINR level and lower connection outage probability can be achieved. For the edge zone case in Fig. 18, we consider an MUE 420 m from the MBS, which is near the boundary of the edge zone. Our proposed scheme again provides lower connection outage probability even though the SINR threshold value is increased. This is also because of the low power transmission of highly interfering frequency bands.

Performance comparison for MUE with respect to traffic load are shown in Figs. 19-21. We assume the ratio of traffic loads in macrocell 1, 2, 3, 4, 5, 6, 7 as 15.5:15:14:13:11:10:21.5 and 16:15:13:12:9:7:28 for set 1 and set 2 traffic intensities, respectively. We define the traffic load as the ratio of required bandwidth and initially allocated bandwidth considering the reused frequency bands only e.g., for macrocell 7, $\textit{traffic load} = (\textit{required bandwith} \notin Z)/C$. We assume average cell dwell time of 120 sec. Fig. 19 shows that the proposed scheme significantly reduces the macrocell user call-blocking probability. The call blocking probability also depends on the traffic intensities in other macrocells. This is due to the fact that the amount of frequency borrowing is related to the traffic load of the macrocells. Fig. 20 shows the comparison of SINR for a MUE at center zone. Set 1 traffic intensity was assumed for the SINR performance comparison. We consider a MUE at a distance of 240 m from the MBS. It shows that the proposed scheme prioritizes the RT traffic calls and provides better level of SINR. Even though, the traffic load is increased, the SINR performance maintains almost equal. Hence, due to increase of traffic load, the call-blocking rate is just increased.

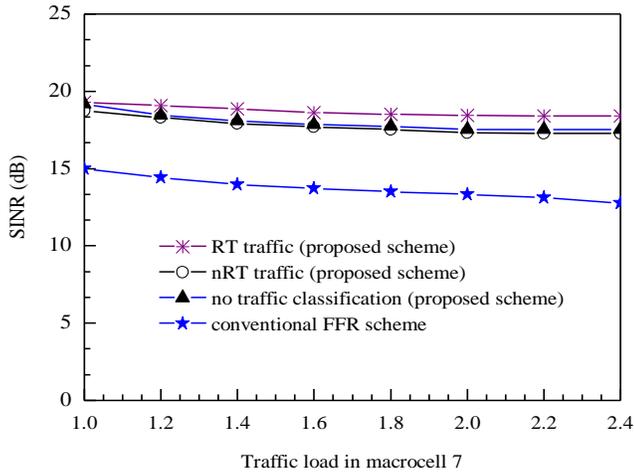

**Fig. 21.** Comparison of SINR for a MUE at edge zone (420 m distance).

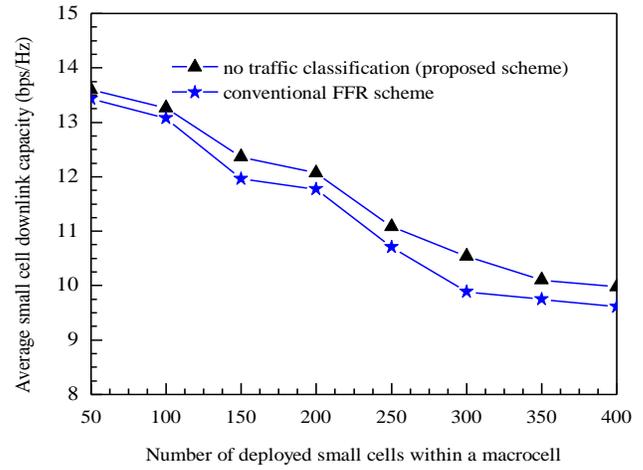

**Fig. 22.** Comparison of small cell downlink capacities for an sUE in the center zone.

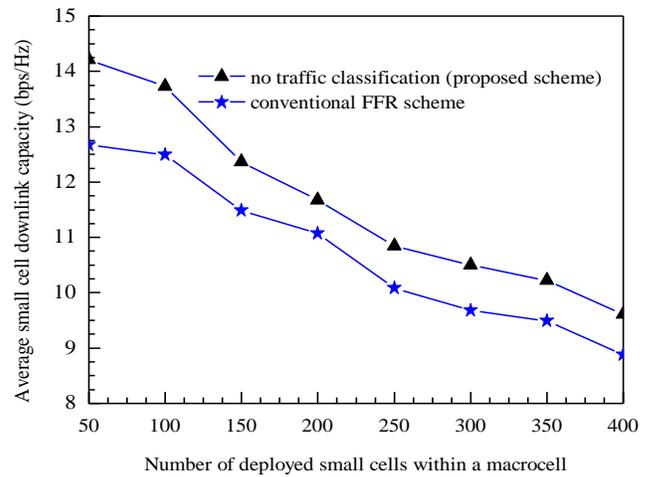

**Fig. 23.** Comparison of small cell downlink capacities for an sUE in the edge zone.

Fig. 21 shows the comparison of SINR for a MUE at macrocell edge zone. We consider a MUE at a distance of 420 m from the MBS. The SINR performance for this case is lower than that of center zone case. This is because the received interference is higher at this zone. It also shows that the SINR performance slightly decreases with the increase of traffic load. Whenever traffic load increases, the probability to use the higher interfering frequency bands also increases. Hence, SINR is decreased. However, the proposed scheme also prioritizes the RT traffic calls and provides better level of SINR.

We also show the performance improvement on sUE for our proposed scheme in Figs. 22-25. The performances are evaluated for different number of deployed small cells. The distance between sBS and sUE is considered 8 m. For the center zone of macrocell case, we assume a distance of 200 m between the MBS and the reference sBS whereas we assume 400 m for edge zone case. The results show that, the performance parameters of sUE are not degraded due to the proposed scheme. Moreover, they are improved due to the proper power allocation. Fig. 22 shows a comparison of small cell downlink capacities for an sUE in the center zone. The capacity of the proposed scheme is better than that of the conventional FFR scheme. Fig. 23 shows the comparison for an sUE in macrocell edge zone. The proposed scheme improves the capacity significantly. Due to the proper power allocation, a significant reduced interference level is achieved from the neighbor macrocells. As the number of deployed small cells is increased, the inter-small-cell interference is increased and therefore, the achievable average capacity is decreased.

Figs. 24 and 25 show that our proposed scheme also keeps the connection outage probability for an sUE within the acceptable range even though a high number of sBS are deployed within a macrocell. The sUE may receive interference signals from all of the neighbor macrocells as well as few neighbor small cells. Fig. 24 shows a comparison of connection outage probabilities for an sUE in the center zone. Due to proper power allocation for reassigned frequency bands, it is possible to achieve interference level reduction significantly from the neighbor macrocells and finally, the connection outage probability is minimized. Fig. 25 shows the performance improvement for an sUE in macrocell edge zone.

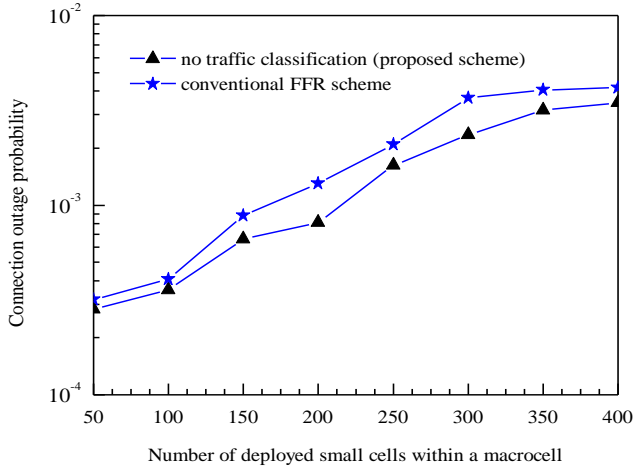

**Fig. 24.** Comparison of connection outage probabilities for an sUE in the center zone.

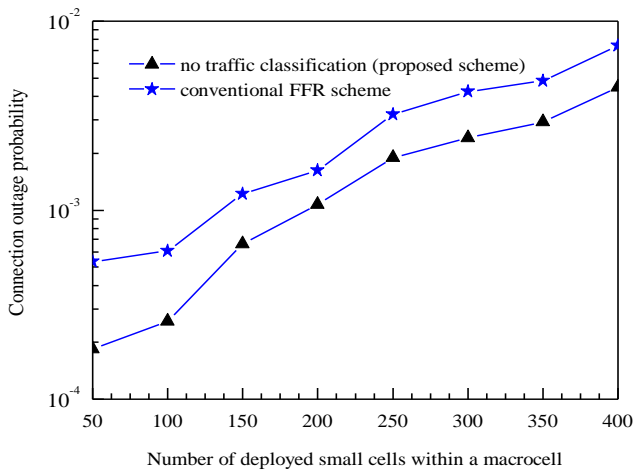

**Fig. 25.** Comparison of connection outage probabilities for an sUE in the edge zone.

## VI. CONCLUSIONS

Cellular networks always have less resources than required to support the growing demand of various high data rate services. Therefore, resource management is an important issue for cellular networks to maximize resources as well as the QoS level. FFR is an excellent approach for increasing spectrum utilization and reducing interference in OFDMA networks. This paper provides a new FFR scheme that considers the case of coexisting small cells and macrocells. In our proposed scheme, the total frequency-band allocations for different macrocells are decided based on the traffic intensity. The scheme also controls the transmitted power to reduce interference effects. The level of transmitted power for different frequency bands is based on the level of interference from other frequency bands. Use of traffic classification in terms of RT and nRT to allocate the frequency bands ensures the maximum QoS level for the higher-priority RT traffic calls without reducing the QoS level of nRT traffic calls. Performance analyses show that our proposed scheme can significantly improve the performance of the conventional FFR technique. We expect that this new concept of FFR will contribute a lot for maximizing the resource utilization in HetNets. As a future research direction, an optimization of the network performance to frequency and power allocation using stochastic geometry will be investigated.

## References

[1] P. D. Lorenzo, S, Barbarossa, and A. H. Sayed, "Distributed spectrum estimation for small cell networks based on sparse diffusion adaptation," *IEEE Signal Processing Letters*, vol. 20, no. 12, pp. 1261-1265, Dec. 2013.
[2] Q.-D. Vu, L.-N. Tran, R. Farrell, and E.-K. Hong, "Energy-efficient zero-forcing precoding design for small-cell networks," *IEEE Transactions on Communications,* vol. 64, no. 2, pp. 790-804, Feb. 2016.
[3] C. Wang, C. Liang, F. R. Yu, Q. Chen, and L. Tang, "Computation offloading and resource allocation in wireless cellular networks with mobile edge computing," *IEEE Transactions on Wireless Communications,* vol. 16, no. 8, pp. 4924-4938, Aug. 2017.
[4] M. Z. Chowdhury and Y. M. Jang, "Handover management in high-dense femtocellular networks," *EURASIP Journal on Wireless Communications and Networking,* pp. 1-21, Jan. 2013.
[5] A.S.M. Z. Shifat, M. Z. Chowdhury, and Y. M. Jang, "Game-based approach for QoS provisioning and interference management in 5G heterogeneous networks," *IEEE Access,* vol. pp, no. 99, pp. 1-1, 2017.
[6] H.-B. Chang and I. Rubin, "Optimal downlink and uplink fractional frequency reuse in cellular wireless networks," *IEEE Transactions on Vehicular Technology*, vol. 65, no. 4, pp. 2295-2308, Apr. 2016.
[7] S. Kumar, S. Kalyani, K. Giridhar, "Optimal design parameters for coverage probability in fractional frequency reuse and soft frequency reuse," *IET Communications*, vol. 9, no. 10, pp. 1324-1331, Jun. 2015.
[8] A. S. Mohamed, M. Abd-Elnaby, and S. A. El-Dolil, "Self-organised dynamic resource allocation scheme using enhanced fractional frequency reuse in long term evolution-advanced relay-based networks," *IET Communications*, vol. 10, no. 10, pp. 1163-1174, Jul. 2016.
[9] N. Al-Falahy and O. Y. K. Alani, "Network capacity optimisation in millimetre wave band using fractional frequency reuse," *IEEE Access,* vol. PP, no. 99, pp. 1-1, 2017.
[10] H. Tabassum, Z. Dawy, M. Sl. Alouini, and F. Yilmaz, "A generic interference model for uplink OFDMA networks with fractional frequency reuse," *IEEE Transactions on vehicular technology*, vol. 63, no. 3, pp. 1491-1497, March 2014.
[11] Q. Li, R. Q. Hu, Y. Xu, and Y. Qian, "Optimal fractional frequency reuse and power control in the heterogeneous wireless networks," *IEEE Transactions on Wireless Communications,* vol. 12, no. 6, pp. 2658-2668, Jun. 2017.
[12] M. M. Aldosari and K. A. Hamdi, "Energy efficiency of distributed antenna systems using fractional frequency reuse," *IEEE Communications Letters*, vol. 19, no. 11, pp. 1985-1988, Nov. 2015.
[13] M. Z. Chowdhury, M. A. Hossain, S. Ahmed, and Y. M. Jang, "Radio resource management based on reused frequency allocation for dynamic channel borrowing scheme in wireless networks," *Wireless Networks,* vol. 21, no.8, pp. 2593-2607, Nov. 2015.
[14] C. C. Coskun, K. Davaslioglu, and E. Ayanoglu, "Three-stage resource allocation algorithm for energy-efficient heterogeneous networks," *IEEE Transactions on vehicular technology*, vol. 66, no. 8, pp. 6942-6957, Aug. 2017.
[15] J. G.-Morales, G. Femenias, and F. R.-Palouo, "Analysis and optimization of FFR-aided OFDMA-based heterogeneous cellular networks," *IEEE Access*, vol. 6, pp. 5111-5127, 2016.
[16] G. Aliu, M. Mehta, M. A. Imran, A. Karandikar, and B. Evans, "A new cellular-automata-based fractional frequency reuse scheme," *IEEE Transactions on Vehicular Technology*, vol. 64, no. 4, pp. 1535-1547, Apr. 2015.
[17] Z. Xu, G. Y. Li, C, Yang, and X. Zhu, "Throughput and optimal threshold for FFR schemes in OFDMA cellular networks," *IEEE Transactions on Wireless Communications,* vol. 11, no. 8, pp. 2776-2785, Aug. 2012.
[18] A. Mahmud, and K. A. Hamdi, "A Unified framework for the analysis of fractional frequency reuse techniques," *IEEE Transactions on Communications,* vol. 62, no. 10, pp. 3692-3705, Oct. 2014.
[19] J. Y. Lee, S. J. Bae, Y. M. Kwon, and M. Y. Chung, "Interference analysis for femtocell deployment in OFDMA systems based on fractional frequency reuse," *IEEE Communications Letters*, vol. 15, no. 4, pp. 425-427, Apr. 2011.

[20] J. G.-Morales, G. Femenias, and F. R.-Palou, "Statistical analysis and optimization of FFR/SFR-aided OFDMA-based multi-cellular networks," in Proc. of *IEEE Statistical Signal Processing Workshop (SSP)*, pp. 1-5, Jun. 2016.
[21] C. Rose, S. Ulukus, and R. D. Yates, "Wireless systems and interference avoidance," *IEEE Transactions on Wireless Communications,* vol. 1, no. 3, pp. 415-428, Jul. 2002.
[22] A. S. Hamza, S. S. Khalifa, H. S. Hamza, and K. Elsayed, "A survey on inter-cell interference coordination techniques in OFDMA-based cellular networks," *IEEE Communications Survey & Tutarials,* vol. 15, no. 4, pp. 1642-1670, 2013.
[23] A. Hernandez, I. Guıo, and A. Valdovinos, "Radio resource allocation for interference management in mobile broadband OFDMA based networks," *Wireless Communications and Mobile Computing*, vo.10, no. 11, pp. 1409-1430, Nov. 2010.
[24] IEEE *Part 16: Air Interference for Broadband Wireless Access System Amendment 3: Advanced Air Interface*, IEEE Std. 802.16m-2011, May 2011.
[25] "UTRAN architecture for 3 G Home Node B (HNB); Stage 2," Sophia-Antipolis, France, TS25.467 (Rel. 10), 2011.
[26] Y. Liu, C. S. Chen, C. W. Sung, and C. Singh, "A game theoretic distributed algorithm for FeICIC optimization in LTE-A HetNets," *IEEE/ACM Transactions on Networking*, vol. 25, no. 6, pp. 3500-3513, Dec. 2017
[27] H. Zhou, Y. Ji, X. Wang, and S. Yamada, "eICIC configuration algorithm with service scalability in heterogeneous cellular networks," *IEEE/ACM Transactions on Networking*, vol. 25, no. 1, pp. 520-535, Feb. 2017.
[28] K. I. Pedersen, B. Soret, S. Barcos Sanchez, G. Pocovi and H. Wang, "Dynamic enhanced intercell interference coordination for realistic networks," *IEEE Transactions on Vehicular Technology*, vol. 65, no. 7, pp. 5551-5562, July 2016.
[29] L. Tang, Y. Wei, W. Chen, and Q. Chen, "Delay-aware dynamic resource allocation and ABS configuration algorithm in HetNets bsed on Lyapunov optimization," *IEEE Access*, vol. 5, pp. 23764-23775, 2017.
[30] C. Liu, M. Li, S. V. Hanly and P. Whiting, "Joint downlink user association and interference management in two-tier Hetnets with dynamic resource partitioning," *IEEE Transactions on Vehicular Technology*, vol. 66, no. 2, pp. 1365-1378, Feb. 2017.
[31] M. Z. Chowdhury, Y. M. Jang, and Z. J. Haas, "Call admission control based on adaptive bandwidth allocation for wireless networks," *Journal of Communications and Networks (JCN)*, vol. 15, no. 1, pp. 15-24, Feb. 2013.
[32] M. Z. Chowdhury, Y. M. Jang, and Z. J. Haas, "Cost effective frequency planning for capacity enhancement of femtocellular networks," *Wireless Personal Communications*, vol. 60, no. 1, pp. 83-104, September 2011.
[33] M. Z. Chowdhury and Y. Min Jang, "Class-based service connectivity using multi-level bandwidth adaptation model in multimedia wireless networks," *Wireless Personal Communications*, vol. 77, no 4, pp. 2735-2745, Aug. 2014.
[34] L. Wang, X. Zhang, S. Wang, and J. Yang, "An online strategy of adaptive traffic offloading and bandwidth allocation for green M2M communications," *IEEE Access*, vol. 5, pp. 6444-6453, May 2017.
[35] G. Huang, Z. Lin, D. Tang, and J. Qin, "QoS-driven jointly optimal power and bandwidth allocation for heterogeneous wireless networks," *Electronics Letters*, vol. 51, no 1, pp. 122-124, Jan. 2015.
[36] J. Miao, Z. Hu, K. Yang, C. Wang, and H. Tian, "Joint power and bandwidth allocation algorithm with QoS support in heterogeneous wireless networks," *IEEE Communications Letters*, vol. 16, no. 4, pp. 479-481, Apr. 2012
[37] I. Shgluof, M. Ismail, and R. Nordin, "Semi-Clustering of victim-cells approach for interference management in ultra-dense femtocell networks," *IEEE Access*, vol. 5, pp. 9032-9043, May 2017.
[38] A. Mukherjee, D. De, and P. Deb, "Interference management in macro-femtocell and micro-femtocell cluster-based long-term evaluation-advanced green mobile network," *IET Communications*, vol. 10, no 5, pp. 468-478, Apr. 2016.
[39] H. Wang, C. Zhu, and Z. Ding, "Femtocell power control for interference management based on macrolayer feedback," *IEEE Transactions on Vehicular Technology*, vol. 65, no. 7, pp. 5222-52236, July 2016.
[40] K. Pahlavan and P. Krishnamurthy, *Principles of Wireless Networks*, Prentice Hall PTR, New Jersey, 2002.
[41] Femtoforum, "OFDMA Interference Study: Evaluation Methodology Document," pp. 9-15, November 2008.